\documentclass[english]{iopart}

\usepackage{graphicx}
\usepackage{url}
\usepackage{iopams} 

\expandafter\let\csname equation*\endcsname\relax
\expandafter\let\csname endequation*\endcsname\relax

\usepackage{amssymb}
\usepackage[latin1]{inputenc}
\usepackage{amsmath}
\usepackage{epsfig}

\newcounter{fig}
\begin{document}

\title[Susceptibility of the Ising model mod. prime]
{\Large Automata and the susceptibility of the square lattice Ising model modulo powers of primes.}

\vskip .3cm 

\author{A. J. Guttmann$^\ddag$, 
J-M. Maillard$^\dag$}
\address{$^\ddag$ School of Mathematics and Statistics,
The University of Melbourne, Victoria 3010, Australia
}
\address{$^\dag$ LPTMC, UMR 7600 CNRS, 
Universit\'e de Paris 6, Sorbonne Universit\'es, 
Tour 23, 5\`eme \'etage, case 121, 
 4 Place Jussieu, 75252 Paris Cedex 05, France} 

\vskip .2cm 

Email: T.Guttmann@ms.unimelb.edu.au   and   maillard@lptmc.jussieu.fr

\vskip .2cm 

\vskip .3cm 

{\em Dedicated to R.J. Baxter, for his 75th birthday.}

\vskip .3cm 


\begin{abstract}

We study the full susceptibility of the Ising model modulo powers of primes. 
We find exact functional equations for the full susceptibility modulo these
primes. Revisiting some lesser-known results on discrete finite automata,
we show that these results can be seen as a consequence of the fact
that, modulo $\, 2^r$, one cannot distinguish the full susceptibility
from some simple diagonals of rational functions which reduce to 
algebraic functions modulo $\, 2^r$, and, consequently, 
satisfy exact functional equations modulo $\, 2^r$.
We sketch a possible physical interpretation of these functional equations
modulo $\, 2^r$ as reductions of a master functional equation corresponding
to infinite order symmetries such as the isogenies of elliptic curves. One 
relevant example is the Landen transformation which can be seen as an exact 
generator of the Ising model renormalization group. 
We underline the importance of studying a new class of functions corresponding 
to ratios of diagonals of rational functions: they reduce to algebraic functions
modulo powers of primes and they may have solutions with natural boundaries.

\end{abstract}

\noindent {\bf PACS}: 05.50.+q, 05.10.-a, 02.30.Gp, 02.30.Hq, 02.30.Ik 

\vskip .1cm 

\noindent {\bf AMS Classification scheme numbers}: 03D05, 11Yxx, 33Cxx,  34Lxx,  34Mxx, 34M55,  39-04, 68Q70

\vskip .2cm

 {\bf Key-words}:   susceptibility of the Ising model, modulo prime calculation, 
functional equations, diagonals of rational functions, automaton, 
algebraic power series,  lacunary series, Landen transformations, Frobenius operator.

\vskip .3cm

\section{Introduction}
\label{introduction}

Despite the enormous progress made over the last 75 years in the study of
(Yang-Baxter) integrable models in lattice statistical mechanics and
enumerative combinatorics, there still remain many important unanswered questions.

One of the most intriguing is the susceptibility of the two-dimensional
Ising model.
The closed form expression\footnote[1]{It can be rewritten in 
a simpler $\, _4F_3$ hypergeometric form, see~\cite{Viswanathan}.} of the partition 
function was obtained by L. Onsager~\cite{Onsager} in 1944, and the spontaneous 
magnetization was obtained a few years later by both Onsager (unpublished) 
and Yang~\cite{Yang}. However, after more than 70 years, a closed form expression 
for the  full susceptibility still eludes us. Accordingly,  understanding the 
nature of this function remains a challenging problem. 

Forty years ago, Wu, Barouch, McCoy and Tracy~\cite{wu-mc-tr-ba-76} showed 
that the full susceptibility of the square-lattice
Ising model can be decomposed as the infinite sum 
 of {\em holonomic $\,n$-fold
integrals}~\cite{bo-bo-ha-ma-we-ze-09,Wu,ze-bo-ha-ma-04,ze-bo-ha-ma-05c,Khi6}, 
denoted $\, \chi^{(n)}$. In the last 
decade the linear differential operators 
corresponding to the first  $\, \chi^{(n)}$'s, up to $\, n\, =\, 6$, were obtained, 
underlying the role of the elliptic curve parametrization~\cite{CalabiYauIsing}, but 
showing also the emergence of (at least) one Calabi-Yau ODE, and beyond, of linear 
differential operators with selected differential Galois 
groups~\cite{Special,bridged,Canonical}. A complete description
of the singular points of the linear differential operators
 corresponding to the first few $\, \chi^{(n)}$'s has also been 
obtained~\cite{Wu,Landau,Singularities,bo-gu-ha-je-ma-ni-ze-08}. Despite 
being an infinite sum of 
holonomic $\,n$-fold integrals, the full susceptibility 
is {\em not a holonomic function}~\cite{Orrick}.

Further, in a recent paper it has been shown that these $\,n$-fold
integrals are actually {\em diagonals of rational functions}~\cite{Short,Big}. 
Consequently their series expansions are such that modulo any 
prime, or power of a prime,
they can be identified with the 
{\em series expansions of an algebraic function}~\cite{Short,Big}. These 
properties were explicitly shown
in the case of $\, \chi^{(3)}$, in section (3.1) of~\cite{Short}. In 
particular it was shown that $\, \, H(w) \, = \, \,  {\tilde{\chi}}^{(3)}/8$
(defined in Section \ref{definition}, see (\ref{tildechi3})), satisfies, 
modulo $\, 2$, the quadratic equation 
\begin{eqnarray}
\label{degree2}
\hspace{-0.9in}&&\qquad \qquad \quad 
H(w)^2\, \,  + w \cdot \, H(w) \,\,\,   + w^{10}
  \,   \, \, \,  \, = \, \, \, \, \,  \,   0  \quad \quad \mod \,  2.  
\end{eqnarray}
and  modulo $\, 3$, the polynomial equation of degree nine
\begin{eqnarray}
\label{degree9}
\hspace{-0.95in}&&\, 
p_9 \cdot \, H(w)^9 \, 
 + \,  w^{6} \cdot \, p_3 \cdot \, H(w)^3  \, 
+ \,   w^{10} \cdot \, p_1 \cdot \, H(w) \,
+ \, \, w^{19} \cdot \, p_0^{(1)}\cdot \, p_0^{(2)}\, 
 \, \,  = \, \,\, 0, 
\end{eqnarray}
where:
\begin{eqnarray}
\hspace{-0.95in}&&\quad p_0^{(1)} \, \, = \, \, \,
{w}^{6}+{w}^{5}+{w}^{4}-{w}^{2}-w+1,
\nonumber   \\
\hspace{-0.95in}&&\quad p_0^{(2)} \,  = \,  \,{w}^{37} \, - {w}^{36} \, 
+{w}^{35} \, - {w}^{33} \, +{w}^{31} \, -{w}^{30} \, +{w}^{28} \,
 +{w}^{27} \, +{w}^{24} - {w}^{23} +{w}^{22}
 \nonumber  \\
\hspace{-0.95in}&&\qquad \quad  \quad  \quad   \, 
- {w}^{21} - {w}^{18} - {w}^{16}+{w}^{14} \, - {w}^{12} \, 
-{w}^{11} \, -{w}^{10}\, +{w}^{7} \, -{w}^{5} \, -{w}^{3}\, -1, 
\nonumber  \\
\hspace{-0.95in}&&\quad  p_1 \, \, = \, \, \,
 ({w}^{2}+1)^{20} \, (1-w)^{13}, \, \quad \, \quad \quad 
 p_3 \, \, = \, \, \,
 ({w}^{2}+1)^{18} \, (1-w)^{15} \, ({w}^{4}-{w}^{2}-1),
\nonumber  \\
\hspace{-0.95in}&&\quad p_9 \, \, = \, \, \,\, 
  \, (w+1) ^{3} \, ({w}^{2}+1)^{18} \, (w-1)^{24}. 
\end{eqnarray}
Since all the $\, \chi^{(n)}$'s are diagonals of rational functions~\cite{Short}, 
similar results are expected for any  $\, \chi^{(n)}$
modulo any prime $\, p$, and, beyond, modulo any power of a
prime $\, p^r$.
As a consequence of the Fermat relations, $\, a^p \, = \, a$, 
modulo $\, p$, one can expect relations, like (\ref{degree2})
or (\ref{degree9}), to be expressible as functional equations
where $\, H(w)^p$ is replaced by $\, H(w^p)$. 
Now, the full susceptibility $\, \chi$ is not the diagonal of a rational function, 
indeed it is {\em not even holonomic}~\cite{Singularities,Orrick}. Therefore,
 for the full susceptibility, one cannot expect relations like (\ref{degree2})
or (\ref{degree9}) to exist.
Due to the complexity of this  
function\footnote[2]{Which has, for instance, a natural 
boundary~\cite{Singularities,bo-gu-ha-je-ma-ni-ze-08,Orrick}.}, one 
might not expect, at first sight, such functional equations 
for the full susceptibility.

However, as we show below, the full susceptibility, when expressed 
in the appropriate expansion variable, does satisfy some surprisingly 
simple functional equations {\em modulo} certain primes, or power of primes.

These exact results show that the full susceptibility 
{\em reduces to an algebraic function, modulo certain primes, or powers of primes},
and thus sheds new light on the integrable character of this very important function 
in physics. We consider this a surprising result:
we certainly did not expect such simple results for the full susceptibility. 
This gives us considerable incentive to systematically 
study other non-holonomic physical series modulo primes or powers of primes.
It will be interesting to see whether this is an exceptional result, in which case 
it sheds more light on the susceptibility, or a
common occurrence, in which case we need to explain why.

\vskip .2cm 

\section{Definitions and some known results on the full susceptibility.}
\label{definition}

\vskip .1cm 

In 1976, Wu, McCoy, Tracy and Barouch~\cite{wu-mc-tr-ba-76}
 showed that the susceptibility could be expressed as an infinite 
sum of contributions, known as 
{\em $\, n$-particle contributions} $\chi^{(n)}$. The 
low-temperature series were given by the case $n$ even, and the 
high-temperature series by $n$ odd. 
More precisely the low temperature susceptibility is given 
by~\cite{bo-gu-ha-je-ma-ni-ze-08}
\begin{eqnarray}
\label{chimoins}
\hspace{-0.95in}&&   \quad 
kT \cdot \chi_L(w)\, \, \,  = 
 \, \,\, \, (1 - 1/s^4)^{\frac{1}{4}} 
\cdot  \tilde{\chi}_L(w)\, \, \,  = \,  \,
 \, \,\, \, (1 - 1/s^4)^{\frac{1}{4}} 
\cdot \, \sum \tilde{\chi}^{(2n)}(w)
\nonumber \\
\hspace{-0.95in}&&   \quad  \quad  \quad  \quad \quad  \quad \quad 
\, \,  =  \, 
 \, \,\, \, (1 - s_L^4)^{\frac{1}{4}} 
\cdot \sum \tilde{\chi}^{(2n)}(w), 
\end{eqnarray}
in terms of the self-dual temperature variable 
$\, w= \, \frac{1}{2}s/(1+s^2)$, where  $ \,s \,= \, \, \sinh(2J/kT)$, 
and $ \,s_L \,= \, \, 1/\sinh(2J/kT)$. 
The  high temperature susceptibility is given by~\cite{bo-gu-ha-je-ma-ni-ze-08}
\begin{eqnarray}
\label{plus}
\hspace{-0.95in}&&   \quad 
kT \cdot \chi_H(w) \, \,\,  =  \, \, \, \,
{{1} \over {s}} \cdot (1 - s^4)^{\frac{1}{4}}
\cdot \,  \tilde{\chi}_H(w)
 \, \,\,  =  \, \, \, \,
{{1} \over {s}} \cdot (1 - s^4)^{\frac{1}{4}}
 \cdot \, \sum \tilde{\chi}^{(2n+1)}(w).
\end{eqnarray}

Remarkably long series expansions with respectively $\, 2042$ and  
$\, 2043$ coefficients\footnote[1]{The low temperature series 
$\,\tilde{\chi}_L(w),$ being an even function, means that the expansion
is known up to the coefficient of $\, w^{4086}$ (see (\ref{tildechilow})).
}, have been obtained\footnote[2]{Using an algorithm adapted from
the Fortran algorithm in~\cite{Orrick}.}~\cite{Iwan} 
for  $\,\tilde{\chi}_L(w)$ and $\,\tilde{\chi}_H(w)$, namely 
\begin{eqnarray}
\label{tildechilow}
\hspace{-0.95in}&&
\tilde{\chi}_L(w) \,\, = \, \, \, 
 \, \,4\,{w}^{4} \, +80\,{w}^{6}\,+1400\,{w}^{8}
\,+23520\,{w}^{10}\,+388080\,{w}^{12}\,+6342336\,{w}^{14}
\nonumber  \\
\hspace{-0.95in}&&  
 \, \, 
+103062976\,{w}^{16} +1668639424\,{w}^{18} +26948549680\,{w}^{20} 
 + \cdots + \tilde{c}^{(L)}_{4086}\, w^{4086}
 +\,  \cdots  
\end{eqnarray}
and 
\begin{eqnarray}
\label{tildechi}
\hspace{-0.95in}&&   \, \,
\tilde{\chi}_H(w) \, \,\, = \, \, \,
 2\,w\, +8\,{w}^{2}\,+32\,{w}^{3}\,+128\,{w}^{4}\,+512\,{w}^{5}\,
+2048\,{w}^{6}\,+8192 \,{w}^{7}\,+32768\,{w}^{8}
\nonumber \\
\hspace{-0.95in}&&   \quad   \,      \quad \,
\,+131080\,{w}^{9}\,+524288\,{w}^{10}\,
+2097440\,{w}^{11}
\, + \, \cdots \, +\,\tilde{c}^{(H)}_{2043}\, w^{2043}\, \, + \,\, \cdots 
\end{eqnarray}

It is worth comparing these two series with the series corresponding to 
the first $\, \tilde{\chi}^{(n)}(w)$ in the two infinite sums 
(\ref{chimoins}) and (\ref{plus}), namely :
\begin{eqnarray}
\label{tildechilow2}
\hspace{-0.95in}&&   \quad \quad \quad  
\tilde{\chi}_L^{(2)}(w) \, = \, \, \,
 4\, w^4 \cdot \,
 _2F_1\Bigl([{{3} \over{2}}, \,{{5} \over{2}}],[3], \, 16 \, w^2  \Bigr)   
\nonumber \\
\hspace{-0.95in}&&   \quad   \quad \quad  \quad \, \, \, \,
 \,\, = \, \, \, 4\,{w}^{4} \, \, +80\,{w}^{6} \,\,  +1400\,{w}^{8} \,\,
  +23520\,{w}^{10} \, \, +388080\,{w}^{12} \,\,  +6342336\,{w}^{14} 
\nonumber \\
\hspace{-0.95in}&&   \quad   \quad  \quad \quad  \quad \,\,\,\,\, \,\, 
+103062960\,{w}^{16}  \, +1668638400\,{w}^{18} \, 
+26948510160\,{w}^{20} \,  \, \, +  \,  \, \cdots 
\end{eqnarray}
and 
\begin{eqnarray}
\label{tildechi1}
\hspace{-0.95in}&&   \, \,  
\tilde{\chi}_H^{(1)}(w) \,\, = \, \, \, {{ 2\,w} \over {1\,-4\,w }} 
  \,\, = \, \,  \, \,  2\,w\, \, +8\,{w}^{2}\, \, +32\,{w}^{3}\, \, 
+128\,{w}^{4}\, \, +512\,{w}^{5}\, \,  +2048\,{w}^{6}\,+8192\,{w}^{7}
\nonumber  \\
\hspace{-0.95in}&&   \quad  \quad  \quad   \quad  \,\,\,
\,\,+32768\,{w}^{8}\, +131072\,{w}^{9}\, 
+524288\,{w}^{10}\, + \, 2097152\,{w}^{11}\,
 \, + \,\, \cdots
\end{eqnarray}
It is known that $\tilde{\chi}^{(n)} = O(w^{n^2}),$ so that 
the coefficients are the same up to $\, w^{14}$ for the 
low-temperature series, and up to $\, w^{8}$ for the high-temperature 
series. Further, one observes
 that the ratio of the coefficients for $\, \tilde{\chi}_L$ 
and $\, \tilde{\chi}_L^{(2)}$ (resp. $\,\tilde{\chi}_H$ and 
$\,\tilde{\chi}_H^{(1)}$) is very close to $\, 1$.

The series expansion for $\, \tilde{\chi}_L^{(4)}(w)$ reads\footnote[1]{Since 
there is an overall integer of the form $\, 2^{n}$ for all the coefficients 
of the $\, \tilde{\chi}_L^{(n)}(w)$ or $\, \tilde{\chi}_H^{(n)}(w)$
series, we divide them, in the following, 
by an appropriate power of $\, 2^{n}$ factor. The series expansion 
remains an expansion with (smaller) integer coefficients. 
} 
\begin{eqnarray}
\label{tildechi4}
\hspace{-0.95in}&&    
{{\tilde{\chi}_L^{(4)}(w)} \over {2^4}}  \, \, \, = \, \, \,  \, \,
{w}^{16} \, +64\,{w}^{18} \, +2470\,{w}^{20} \, +74724\,{w}^{22} \,
 +1954688\,{w}^{24} \, +46428552\,{w}^{26} 
\nonumber  \\
\hspace{-0.95in}&&   \quad     \,  \, \, \,+1029903288\,{w}^{28}\,
+21716367896\,{w}^{30}\,+440440693418\,{w}^{32}\,
+8663350828976\,{w}^{34}
\nonumber  \\
\hspace{-0.95in}&&   \quad    \, 
\,\,\, +166258457615526\,{w}^{36}\,
+3126949985578700\,{w}^{38} \, +57833406662680980\,{w}^{40}
\nonumber  \\
\hspace{-0.95in}&&   \quad     \, 
\,\,\, +1054656431047823680\,{w}^{42}\, +19003412267837223432\,{w}^{44}
\,\,\, + \, \, \cdots
\end{eqnarray}
The series expansion for $\, \tilde{\chi}_L^{(6)}(w)$ reads 
\begin{eqnarray}
\label{tildechi6}
\hspace{-0.95in}&&     \quad 
{{\tilde{\chi}_L^{(6)}(w)} \over {2^6}}  \, \, \, = \, \, \,  \, \,
{w}^{36}\,+144\,{w}^{38}\,+11306\,{w}^{40}\,
+641604\,{w}^{42}\,+29455804\,{w}^{44}
\nonumber  \\
\hspace{-0.95in}&&   \quad     \quad \, \, 
\, \,+1161654484\,{w}^{46}\,+40827303872\,{w}^{48} 
\,+1310513628660\,{w}^{50}\,
\nonumber  \\
\hspace{-0.95in}&&   \quad    \quad \, \, 
\,\,+39090651539936\,{w}^{52}\,+1097452668063296\,{w}^{54}
+29281457807054052\,{w}^{56}
\nonumber  \\
\hspace{-0.95in}&&   \quad   \quad \, \, 
\,\,+748130523334531340\,{w}^{58} \, +18414177309344582452 \,{w}^{60}
\, \,\,+ \,\,\, \cdots 
\end{eqnarray}
The difference between $\, \tilde{\chi}_L(w)$ and $\, \tilde{\chi}_L^{(2)}(w)$ 
reads: 
\begin{eqnarray}
\label{tildechi2diff}
\hspace{-0.95in}&&   \,    \,   \, 
\tilde{\chi}_L \, - \, \, \tilde{\chi}_L^{(2)}
 \,\, \, = \, \, \, 16 \,{w}^{16}\,
+1024 \,{w}^{18}\, +39520 \,{w}^{20}\,  +1195584\,{w}^{22}
+31275008\,{w}^{24}\,
\nonumber  \\
\hspace{-0.95in}&&   \quad         \, \,  \,\, 
+742856832\,{w}^{26}
\,+16478452608\,{w}^{28}\, +347461886336\,{w}^{30}
\,+7047051094688\,{w}^{32}
\nonumber  \\
\hspace{-0.95in}&&   \quad       \, \, \,\,\,
+138613613263616\,{w}^{34}\, +2660135321848480\,{w}^{36} 
\,\,\, \,  + \,\,\, \cdots 
\end{eqnarray}

The difference between $\, \tilde{\chi}_L(w)$ and 
$\, \tilde{\chi}_L^{(2)}(w) \, + \, \tilde{\chi}_L^{(4)}(w)$ 
reads: 
\begin{eqnarray}
\label{tildechi24diff}
\hspace{-0.95in}&&   \,     \quad    
\tilde{\chi}_L \, - \, \, (\tilde{\chi}_L^{(2)} + \, \, \tilde{\chi}_L^{(4)})
 \,\, \, = \, \, \, \,  \, 
64\,{w}^{36} \,+9216\,{w}^{38} \,+723584\,{w}^{40} \,+41062656\,{w}^{42}
\nonumber  \\
\hspace{-0.95in}&&   \quad       \quad \,\, \, 
+1885171456\,{w}^{44} \,+74345886976\,{w}^{46}
 \,+2612947447808\,{w}^{48} \,\,\,  + \, \, \, \cdots 
\end{eqnarray}

The difference between $\, \tilde{\chi}_L(w)$ and 
$\, \tilde{\chi}_L^{(2)}(w) \, + \, \tilde{\chi}_L^{(4)}(w)$
$\, + \, \tilde{\chi}_L^{(6)}(w)$ 
reads: 
\begin{eqnarray}
\label{tildechi246diff}
\hspace{-0.95in}&&   \,   \quad   
\tilde{\chi}_L \, 
- \,  (\tilde{\chi}_L^{(2)} + \,  \tilde{\chi}_L^{(4)}\, + \,  \tilde{\chi}_L^{(6)})
 \,\, \, = \, \, \, \, 256 \,{w}^{64} \,
 +65536\,{w}^{66} \, +8815104\,{w}^{68} 
\nonumber \\
\hspace{-0.95in}&&   \quad   \quad  \, \,  \, \,   \,
 +829038592\,{w}^{70} 
+61219149824\,{w}^{72} \, +3779726083072\,{w}^{74} 
\nonumber \\
\hspace{-0.95in}&&   \quad   \quad \,\,\, \,  \, 
 +202925982372864\,{w}^{76} \, 
+9729999547422720\,{w}^{78} \, +424756293921653248\,{w}^{80}
 \nonumber \\
\hspace{-0.95in}&&   \quad   \quad  \, \,  \, 
 \,\, +17127494149322319872\,{w}^{82} \, 
+645117850681779326976\,{w}^{84}\,\,  \,\, +  \,  \, \cdots 
\end{eqnarray}

\vskip .1cm 

The series expansion for $\, \tilde{\chi}_H^{(3)}(w)$ reads 
\begin{eqnarray}
\label{tildechi3}
\hspace{-0.95in}&&     
{{\tilde{\chi}_H^{(3)}(w)} \over {2^3}}  \, \, \, = \, \, \,  \, \,
{w}^{9} \, +36\,{w}^{11}\, +4\,{w}^{12}\, +884\,{w}^{13}\, 
+196\,{w}^{14}\, +18532\,{w}^{15}\, +6084\,{w}^{16}
\nonumber \\
\hspace{-0.95in}&&   \quad  \quad   \,  \, 
\,\, +357391\,{w}^{17}\, +153484\,{w}^{18}\, +6556516\,{w}^{19}\, 
+3440964 \,{w}^{20}\, +116449960\,{w}^{21}
\nonumber \\
\hspace{-0.95in}&&   \quad   \quad   \,  \, 
\,\, +71553656\,{w}^{22} \, +2022814844\,w^{23}\, 
+1413292572\, w^{24} \, +34583048616\, w^{25}
\nonumber \\
\hspace{-0.95in}&&   \quad  \quad   \,  \, 
 \,\, +26900157072\,{w}^{26} \, 
+584324509812\,{w}^{27} \,+498048104276\,{w}^{28}
\, \, + \,\, \, \cdots 
\end{eqnarray}

The series expansion for $\, \tilde{\chi}_H^{(5)}(w)$ reads 
\begin{eqnarray}
\label{tildechi5}
\hspace{-0.95in}&&    
{{\tilde{\chi}_H^{(5)}(w)} \over { 2^5}}  \, \, \, = \, \, \,  \, \,
{w}^{25} \,+100\,{w}^{27} \,+5652\,{w}^{29} \,+4\,{w}^{30} \,
+238032\,{w}^{31} \,+484\,{w}^{32}  \,+8323743\,{w}^{33} 
 \nonumber \\
\hspace{-0.95in}&&   \quad  \,  \, \, \,\, +32436\,{w}^{34} \,
+255716632\,{w}^{35} \,+1592488\,{w}^{36} \,+7139250236\,{w}^{37}
 \,+63994900\,{w}^{38}
  \nonumber \\
\hspace{-0.95in}&&   \quad \,\, \,  \, \,
+185181953320\,{w}^{39} \,+2231760988\,{w}^{40} 
\,+4531508893397\,{w}^{41} \,
+69986224204\,{w}^{42} \nonumber \\
\hspace{-0.95in}&&  \quad \, \,  \,
\, \,+105775797597812\,{w}^{43} \,+2020409460692\,{w}^{44} \,
+2374723605151320\,{w}^{45} \nonumber \\
\hspace{-0.95in}&&   \quad\, \,  \, \, \,
+54584651129624\,{w}^{46}
 \,+51602310149637388\,{w}^{47} \,
+1396760803374712\,{w}^{48} \nonumber \\
\hspace{-0.95in}&&   \quad \, \, \, 
\, \,+1090696414153653447\,{w}^{49} \,\,  \, \,+  \, \, \, \cdots 
\end{eqnarray}

Comparing  $\,\tilde{\chi}_H(w)$ and the sum 
 $\,\tilde{\chi}_H^{(1)}(w) \, + \, \tilde{\chi}_H^{(3)}(w)$ one finds that 
these two series are the same up to $\, {\rm O}(w^{25}),$ as expected:
\begin{eqnarray}
\label{tildechi3diff}
\hspace{-0.95in}&&  \, \,  \, 
\tilde{\chi}_H \, - \,  (\tilde{\chi}_H^{(1)} \, + \, \tilde{\chi}_H^{(3)})
 \, \,\, = \, \, \,\, 32\,{w}^{25} \, +3200\,{w}^{27}\, +180864\,{w}^{29}\, 
+128\,{w}^{30}\, +7617024\,{w}^{31}
\nonumber \\
\hspace{-0.95in}&&   \quad  \quad  \, \, \, \, 
 +15488\,{w}^{32}\, +266359776\,{w}^{33}
\, +1037952\,{w}^{34} \, +8182932224\,{w}^{35}
\, \,\, + \, \, \, \cdots 
\end{eqnarray}
and
\begin{eqnarray}
\label{tildechi5diff}
\hspace{-0.95in}&&  \quad \quad \quad \,  \, \,
\tilde{\chi}_H \, 
- \,  (\tilde{\chi}_H^{(1)} \, 
+ \, \tilde{\chi}_H^{(3)} \, + \, \tilde{\chi}_H^{(5)})
 \, \,\, = \, \, \,\,  128 \, w^{49} \, + \,25088\,{w}^{51}\,+2621952\,{w}^{53}
 \nonumber \\
\hspace{-0.95in}&&   \quad \quad \quad  \quad \quad  \, 
\,+194185216\,{w}^{55}\,+512\,{w}^{56}\,+11431676800\,{w}^{57}
\,+115200\,{w}^{58}
 \nonumber \\
\hspace{-0.95in}&&   \quad \quad \quad  \quad \quad  \, 
\,+569065324032\,{w}^{59}
\,+13709824\,{w}^{60} \, \,  \, +\, \, \, \cdots 
\end{eqnarray}

\vskip .3cm 

Since the modulus of 
{\em elliptic functions parametrising the Ising model}~\cite{CalabiYauIsing,Singularities} 
is $\, k \, = \, \, s^2$, with the conjectured 
{\em natural boundary}~\cite{bo-gu-ha-je-ma-ni-ze-08}
 corresponding to the unit $\, k$ or $\, s$ circle, it is natural 
to introduce series expansions 
in the $\, s$ or $\, s_L$ variables. In fact, we have studied
 series expansions in the $\, v \, = \, s_L/2\, = \, 1/(2\,s)$
variable in the low-temperature regime, and the $\, v \, = \, s/2$ variable in the 
high-temperature regime, in order to have series with {\em integer coefficients}
(instead of rational coefficients with denominators of 
the form $\, 2^n$). The corresponding low and high temperatures series 
$\, \chi_L(v) \,\, = \, \, \,  \chi_L(s_L/2)$ 
and $\, \chi_H(v) \,\, = \, \, \,  \chi_H(s/2)$ read 
respectively\footnote[1]{Throughout
this paper the $\, {\tilde \chi}$ are functions of the variable $\, w$, while
the  $\, \chi$ are functions of the variable $\, v$.
}
\begin{eqnarray}
\label{chilow}
\hspace{-0.95in}&&  \,  \,
\chi_L(v) \,\, = \, \, \,\,  \, 
4\,{v}^{4} \, \,+16\,{v}^{6}\,\,+104\,{v}^{8}\,\,+416\,{v}^{10}\,\,
+2224\,{v}^{12}\,\,+8896\,{v}^{14} \, +43840\,{v}^{16}
\nonumber \\
\hspace{-0.95in}&&     \,  \quad   \quad 
 \,\,+175296\,{v}^{18}\,+825648\,{v}^{20}\, \,
+3300480\,{v}^{22}  \, +  \,  \cdots   \,+ c^{(L)}_{4086}\, v^{4086}
\,  \,  + \,\, \cdots  
\end{eqnarray}
and  
\begin{eqnarray}
\label{chi}
\hspace{-0.95in}&& 
\chi_H(v)  \,\, = \, \,\, \,
1\,\, +4\,v \,\,+12\,v^2 \,\, +32\,v^3\, \, 
+76\,v^4 \, \,+176\,v^5\, \,+400\,v^6\, \, +896\,v^7\,  +1964\,v^8\
\nonumber \\
\hspace{-0.95in}&&   \quad     \quad \,
 \, +  4256 \,v^9\, 
  \,+9184 \, v^{10} \,+19728 \, v^{11}\, + 41952 \, v^{12} \, +
\, \cdots
 \,+\,c^{(H)}_{2043}\, v^{2043}\, + \,\, \cdots 
\end{eqnarray}
which can be compared with  $\, \chi_L^{(2)} \, = \, \,
 (1 \, -16\, v^4)^{1/4} \cdot \,   \tilde{\chi}_L^{(2)}$
\begin{eqnarray}
\label{chilow2bis}
\hspace{-0.95in}&&   \quad 
\chi_L^{(2)}(v) \,\, = \, \, \,\, 
  {{1} \over {4^3}} \, \cdot \, (1 \, -16\, v^4)^{1/4}
 \cdot \,  \Bigl({{ 4 \, v} \over {1\, +4\, v^2 }}\Bigr)^4 \cdot \,
 _2F_1\Bigl([{{3} \over{2}}, \,{{5} \over{2}}],[3], \,
 \, \Bigl({{4 \, v} \over {1\, +4\, v^2 }}\Bigr)^2   \Bigr)
\nonumber \\
\hspace{-0.95in}&&   \quad \quad   \quad  \quad  
   \,\, = \, \,  \,  \, \,
4\,{v}^{4} \,\, +16\,{v}^{6}\, \,+104\,{v}^{8}\, 
+416\,{v}^{10}\,\, +2224\,{v}^{12}\,\, +8896\,{v}^{14}\, +43824\,{v}^{16}
\nonumber \\
\hspace{-0.95in}&&   \quad \quad   \quad  \quad  \quad   \quad  \quad  \quad  \quad \, \,
  \, +175296\,{v}^{18}\, +825104\,{v}^{20}\, +3300416\,{v}^{22}
 \, \, \,  + \, \, \cdots 
\end{eqnarray}
and 
$\, {\chi}_H^{(1)} \,  = \, \, (1\, -s^4)^{1/4}/s \cdot \, \tilde{\chi}_H^{(1)}$
\begin{eqnarray}
\label{tildechi1ter}
\hspace{-0.95in}&&   \quad 
{\chi}_H^{(1)}(v)  \,\, = \, \, \,
 {{(1\, -s^4)^{1/4}} \over {s}} \cdot \, 
{{s } \over {(1\, -s)^2}}
\,\, = \, \, \, 
 \Bigl({{1\, -16 \,v^4} \over {(1\, -2 \, v)^8}}\Bigr)^{1/4}
\nonumber \\
\hspace{-0.95in}&&   \quad   \quad \,  \,\, \,= \, \, \,\, \,
1\,\,  +4\,v\,\,  +12\,{v}^{2}\,\,  +32\,{v}^{3}\,\,  +76\,{v}^{4}\,\,  
+176\,{v}^{5}\, \,  +400\,{v}^{6}\,\,  +896\,{v}^{7}  \,  +1960\,{v}^{8} 
\nonumber \\
\hspace{-0.95in}&&   \quad   \quad   \quad  \quad  \quad \quad  \, \, \, \,
 \, +4256\,{v}^{9} \,
 \, +9184\,{v}^{10} \,+19712\,{v}^{11} \, +41888 \,\,{v}^{12} \, + \,  \, \cdots
\end{eqnarray}
As must be the case, the coefficients are the same up to $\, v^{14}$ for the 
low-temperature series, and up to $\, v^{8}$ for the high-temperature series, 
and, beyond, the ratio of the coefficients for $\, \chi_L(v)$ 
and $\, \chi_L^{(2)}(v)$ (resp. $\, \chi_H(v)$ and $\, \chi_H^{(1)}(v)$)
are very close to $\, 1$. For the low-temperature series expansion the difference 
between  $\, \chi_L(v)$ and  $\, \chi_L^{(2)}(v)$ reads:
\begin{eqnarray}
\label{tildechidiff2}
\hspace{-0.95in}&&   \,  
\chi_L(v)  - \,  \chi_L^{(2)}(v)
 \,  = \, \,  \,  
16\,{v}^{16} \, +544\,{v}^{20} \, +64\,{v}^{22} \, +13056\,{v}^{24}
\, +2944\,{v}^{26} \, +272512\,{v}^{28}
\nonumber \\
\hspace{-0.95in}&&   \quad   \quad    \quad  \, \, \,    \quad   
 \, \, +88448\,{v}^{30} \, 
+5286560\,{v}^{32} \, +2201856\,{v}^{34} \, +98136096\,{v}^{36} \, 
\,\,  + \, \, \cdots 
\end{eqnarray}

\vskip .1cm

\vskip .1cm
 
It is worth recalling that the very long (low and high temperature) 
series expansions have been obtained for the full susceptibility
as a consequence of a 
{\em quadratic finite difference Painlev\'e functional equation}~\cite{Orrick},
yielding an $\, N^4$ polynomial algorithm for calculating the series. This 
series is therefore ``{\em algorithmically integrable}''. Furthermore 
the $\,n$-fold integrals of the infinite sum 
decomposition~\cite{wu-mc-tr-ba-76}, the $\, \chi^{(n)}$'s,  
have been shown to be highly selected holonomic functions, namely 
{\em diagonals of rational functions}~\cite{Short}.

These properties (``algorithmic integrability'', infinite sums of diagonals of 
rational functions, ...)  suggest that {\em transcendental non-holonomic functions} 
such as the full susceptibility of the square Ising model, should 
correspond to  a ``rather special class'' of {\em non-holonomic} functions,  which 
require new concepts and tools to characterize and analyze them. 
 
\vskip .1cm 

Obtaining such remarkably long series for the full susceptibility was 
a computational ``tour de force,'' and it is likely that these series 
have much more to tell us. To date they have only 
been used to obtain some results on  $\, \chi^{(5)}$ and  $\, \chi^{(6)}$, 
 to confirm exact results~\cite{Landau,Singularities} on the singularities 
of the linear ODEs of the $\, \chi^{(n)}$'s, 
and to clarify the natural boundary scenario~\cite{bo-gu-ha-je-ma-ni-ze-08}. 

In the following sections we revisit these remarkably long series from a new  
{\em finite automaton}~\cite{Poorten} perspective, which in effect means 
considering the various series {\em modulo} various integers, in particular, 
taking a ``$p$-adic'' perspective~\cite{Gouvea}, modulo integers that 
are {\em integer powers of primes}.  
 
\vskip .1cm 

\vskip .1cm 

\section{Functional equations modulo $\, 2^r$ 
for the full susceptibility.}
\label{lowtemp}

\subsection{The low-temperature susceptibility.}
\label{avallowtemp}

Consider the low temperature series (\ref{chilow}) for the full 
susceptibility~\cite{Iwan}, for which $\, 2043$ coefficients have 
been obtained in the $\, u\, = \, \, v^2$ variable~\cite{Iwan}. We 
denote this series $\, F(u),$ so that
\begin{eqnarray}
\label{goodchilow}
\hspace{-0.95in}&&   \quad  
F(u) \,\, = \, \,\, 
4\,{u}^{2} \, +16\,{u}^{3} \, +104\,{u}^{4} \, +416\,{u}^{5} 
\, +2224\,{u}^{6} \, +8896\,{u}^{7}\,  +43840 \,{u}^{8}
\nonumber \\ 
\hspace{-0.95in}&&   \quad  \quad  \quad    \quad    \quad  
\, +175296\,{u}^{9}\, +825104\,{u}^{10}
\, \, + \, \, \cdots \,\,\,  + \, \,a_{2043} \, \cdot u^{2043} 
\, \, \,  + \, \, \cdots \,
\end{eqnarray}
Now consider this series modulo various integers 
$\, q \, = \, \, 2^r$, ($q\, = \, \, 2, \, 4, \, 8,\, 16,$ 
$32, \, 64, \, \cdots$)
where we denote by $\, F_q$ the corresponding series modulo $\, q$. 
We found the following simple results:
\begin{eqnarray}
\label{goodchilowmod}
\hspace{-0.95in}&&   \quad 
F_2(u) \,\, = \, \,\, 0,\, \, \, \quad F_4(u) \,\, = \, \,\, 0, 
\quad \, \,\, 
F_8(u) \,\, = \, \,\, 4 \, u^2, \quad \, \,\, 
 F_{16}(u) \,\, = \, \,\, 4 \, u^2 \, +8\, u^4, \quad 
\end{eqnarray}
where the first two results are of no significance, and just reflect 
the lattice symmetry.
However for $\, q= \, 32$ and $\, q=\, 64$, we found the appearance 
of simple {\em lacunary series}, so that
\begin{eqnarray}
\label{goodchilowmod32}
\hspace{-0.95in}&&   \quad  
 F_{32}(u) \,\, = \, \,\,\,  
20\, u^2 \,\, +24\, u^4 \,\, + \, 16 \cdot \, u^2 \cdot \, L(u), 
 \\
\label{goodchilowmod32bis}
\hspace{-0.95in}&&  \, \quad  
F_{64}(u) \, = \, \, \, 
60\, u^2 \cdot \, (11 +8\,u +10 \, u^2 +8\,u^4 +8\,u^6)
 \,  \, \,+ \,  (48 \, u^2  + \, 32\, u^4) \cdot \, L(u), 
\end{eqnarray}
where $\, L(u)$  corresponds to the first $\, 1024$ coefficients 
of the lacunary series with a natural boundary 
on the unit-circle $\, |u| \, =1$:
\begin{eqnarray}
\label{lacunary}
\hspace{-0.95in}&&  \quad 
L(u) \,\,\, = \, \,\,\, \sum_{n\, = \,0}^{n\, = \,\infty} \, u^{2^n} 
\\
\hspace{-0.95in}&&   \quad   \,  \quad 
\,\, = \, \,\, \,
 1 +u +{u}^{2} +{u}^{4} +{u}^{8} +{u}^{16} +{u}^{32}
+{u}^{64} +{u}^{128} +{u}^{256} +{u}^{512} +{u}^{1024}
\,\,  + \, \cdots
 \nonumber  
\end{eqnarray}
This strongly suggests that $\, F_{32}(u)$ and $\, F_{64}(u)$
{\em  satisfy the modulo $\, 32$ and modulo $\, 64$
 functional equations respectively:} 
\begin{eqnarray}
\label{goodchilowmod32funcequ}
\hspace{-0.95in}&&   \quad  \quad  \quad  \quad   \quad   \quad 
u^2 \cdot \, F_{32}(u)\,\,\,  = \, \,\, \,  \, 
 F_{32}(u^2) \,\,\, \,   +16\,{u}^{5} \,\, +24\,{u}^{6}\, \, +8\,{u}^{8}, 
\end{eqnarray}
\begin{eqnarray}
\label{goodchilowmod64funcequ}
\hspace{-0.95in}&&   \quad  \quad   
u^2 \cdot (3 \, +2 \, u^4) \cdot  \, F_{64}(u)
 \,\,\,   = \,\,  \, \,
 (3 \, +2 \, u^2) \cdot \, F_{64}(u^2) \,
\nonumber \\
\hspace{-0.95in}&&   \quad   \quad   \quad  \quad 
\,\,\,   + \, 16  \, u^5 \cdot \, 
(3 \,+4\,u \, +6\,{u}^{2} \,+2\,{u}^{4} \,
+58 \,{u}^{5} \,+4\,{u}^{6} \,+2\,{u}^{7} \,+6\,{u}^{11}).
\end{eqnarray}
The series expansions $\, F_{128}$, $\, F_{256}$, ... also satisfy similar 
functional equations, but they are more involved, the series having 
a less obvious lacunary series interpretation. For instance 
one finds that 
\begin{eqnarray}
\label{finds}
\hspace{-0.95in}&&  \quad  \quad  
u^2 \cdot \, F_{128}(u)\, \, =\,\,\,   F_{128}(u^2)
 \,\, \, + \,\,\, 
32\, u^6 \cdot \, (3 \, -u^2) \cdot \, L(u)
\,\, + \, 8 \cdot \, u^5 \cdot \, p_{13}, 
\nonumber \\
\hspace{-0.95in}&& \quad  \quad   \hbox{where:}  \, \,  \quad  \quad\quad  
p_{13} \, \, = \, \, \,  2\, -u\, -8\, u^2\, -15\, u^3\, 
+12\, u^4\, -4\, u^5\, 
\nonumber \\ 
\hspace{-0.95in}&& \quad  \quad  \quad  \quad  \quad 
 \quad  \quad   \quad   \quad   \quad  \quad  
+8\, u^6\, +4\, u^7\,
 -8\, u^9\, +4\, u^{11}\, -8\, u^{13},
\end{eqnarray}
where $\, L(v)$ is the lacunary series (\ref{lacunary}),
which satisfies the functional equation 
$\, u \, + \, L(u^2) \, = \, \, L(u)$.
Therefore one deduces the functional equation modulo $\, 128$: 
\begin{eqnarray}
\label{finds}
\hspace{-0.95in}&&     \quad  \,    \quad  
 u^8 \cdot \, (u^4 \, -3) \cdot \, F_{128}(u)\,
- \, u^4 \cdot \, (u^6 \,-2\, u^2 \, -3) \cdot \, F_{128}(u^2)
\nonumber \\
\hspace{-0.95in}&&   \quad    \quad  \quad    \quad    \quad      \quad    
  \,\,\,\, = \,\,\,\,\, \,(u^2 \, -3)  \cdot \, F_{128}(u^4) \, \, + \, \, \, 
16\,\,{u}^{10} \cdot \, p_{28} \quad  \quad  \quad  \quad    \quad \hbox{where:} 
 \\
\hspace{-0.95in}&&  \quad   
p_{28} \, = \, \, 
 4\,{u}^{28}-12\,{u}^{26}-2\,{u}^{24}+6\,{u}^{22}+4\,{u}^{20}
-16\,{u}^{18}+10\,{u}^{14}+10\,{u}^{12}+4\,{u}^{11}
\nonumber \\
\hspace{-0.95in}&&   \quad  \quad  \quad  \, \, \,  
-2\,{u}^{10}+4\,{u}^{9}+12\,{u}^{8}-10\,{u}^{7} -13\,{u}^{6}
-11\,{u}^{5} +11\,{u}^{4} -6\,{u}^{3} -{u}^{2} -3\,u \, +3.
\nonumber 
\end{eqnarray}

\vskip .1cm 

Since we have seen that the full susceptibility series 
is quite close to the series expansion of $\, \chi^{(2)}$, it is natural to ask 
if one obtains similar results modulo $\, 2^r$, for $\, \chi^{(2)}$. From 
the series expansion (\ref{chilow2bis}), we find that 
one obtains {\em the same series as the one displayed in} (\ref{goodchilowmod})
modulo $2, \, 4, \, 8, \, 16.$
Modulo $\, 32$ and $\, 64$ one obtains simple functional equations for $\, \chi^{(2)}$
which are similar  to (\ref{goodchilowmod32funcequ}) and (\ref{goodchilowmod64funcequ}) 
but actually slightly different. 

\vskip .1cm 

This can be rewritten in terms of the difference (\ref{tildechidiff2}).
This difference (\ref{tildechidiff2}) is zero modulo $2, \, 4, \, 8, \, 16$.
 Modulo $\, 32$ it is
just one term, namely $\,16\, v^{16}$ (the series for $\, \chi^{(2)}$ being 
a non-trivial lacunary series) and modulo $\, 64$, it
becomes the lacunary series 
\begin{eqnarray}
\label{tildechidiff2mod64}
\hspace{-0.95in}&&     \quad \quad  
\chi_L(v) \, - \, \chi_L^{(2)}(v)
 \, \,= \,\,\,\,\,
16\,{v}^{16} \,\, +32\,{v}^{20}\,+32\,{v}^{32}\,+32\,{v}^{36}\,
+32\,{v}^{68}\,+32\,{v}^{132}\, 
\nonumber \\
\hspace{-0.95in}&&   \quad  \quad  \quad   \quad  \quad  \quad 
\, +32\,{v}^{260} \,+32\,{v}^{516}\, \, + 32\,{v}^{1028} \, + 32\,{v}^{2052}
\,\,\,\, + \, \,\, \cdots 
\end{eqnarray}

\vskip .1cm 

{\bf Remark:} The low temperature series (\ref{goodchilow})
relied on having coefficients 
up to the term in $\, v^{2043}$. Consequently the previous functional 
equations have been checked up to order $\, 2043$ in the the expansion
(\ref{goodchilow}). The previous calculations
underline the crucial role played by the lacunary
series (\ref{lacunary}) where the next term is
$\,  u^{2048}$. It would thus be interesting to validate a functional equation
such as (\ref{finds}) up to the point where the term $\,  u^{2048}$
in  (\ref{lacunary}) is expected to emerge: this would require one to find 
{\em just\footnote[2]{Even though obtaining more terms for the
low temperature series (\ref{goodchilow}) can be done with a polynomial time
algorithm, getting more coefficients requires substantial computer 
resources: however, here the idea is that we just need 
a few extra terms.
} a few  (less than 10) extra terms} in the low temperature series 
(\ref{goodchilow}). Without trying to get more coefficients for 
the full susceptibility in exact arithmetic (not modulo a prime, 
or a power of a prime) which requires 
very substantial computer ressources, we can try to check all our previous 
functional equations modulo some integers of the form $\, 2^r$, 
seizing the opportunity of having a polynomial algorithm to get
many more than $\, 2000$ coefficients ($5000$, $\,6000$, $\,10000$, ...) but 
just modulo $\, 2$, $\, 4$, $\, 8$, $\, 16$ ...

\vskip .1cm 

\subsection{The high-temperature susceptibility.}
\label{avalhightemp}

Similarly, we now study the high-temperature expansion (\ref{chi}), 
 modulo $ \,q$ with $\, q \, =$
$ \, 2, \, 4, \, 8, \, 16, \, 32, \, 64$, and compare these 
series with the ones obtained modulo $\, q$ with $\, q \, =$ 
$\, \,2, \, 4, \, 8, \, 16, \, 32, \, 64$
 for (\ref{tildechi1ter}). Since, apart from the first constant 
coefficients, all the coefficients are divisible by $\, 4$, we introduce 
the series
\begin{eqnarray}
\label{introduce}
\hspace{-0.95in}&&   \,  
G(v)  \,\, = \, \, \, 
{{\chi(v) \, -1} \over {4}}  \,\, = \, \, \, \, 
v \,\, +3\,{v}^{2} \, +8\,{v}^{3} \, +19\,{v}^{4} \, +44\,{v}^{5} 
\, +100\,{v}^{6} \, +224\,{v}^{7} \, +491\,{v}^{8} 
\nonumber \\
\hspace{-0.95in}&&   \quad   \quad   \quad  \quad  \,  \, 
\, +1064\,{v}^{9} \, +2296\,{v}^{10} \, +4932\,{v}^{11} \, 
+10488\,{v}^{12} \, +22180 \,{v}^{13} \, \, \,  \, +  \,  \, \cdots 
\end{eqnarray}
and denote by $\, G_q$ the corresponding series modulo $\, q$. 
We obtained the following results: Modulo $\, 2$
the series $\, G_2$ is the lacunary series $\, L(v)\, -1$
(where $\, L(v)$ is given by (\ref{lacunary})):
\begin{eqnarray}
\label{Fvv}
\hspace{-0.95in}&&  
G_2(v)  = \,  
v  +{v}^{2}\,+{v}^{4}+{v}^{8}+{v}^{16}+{v}^{32}+{v}^{64}+{v}^{128}
+{v}^{256}+{v}^{512}\, +{v}^{1024} \, + \, \cdots 
\end{eqnarray}
which is a solution of the functional equation 
\begin{eqnarray}
\label{Fvequ}
\hspace{-0.95in}&&   \quad \, \quad  \quad  \quad \quad \quad  \quad  \quad   \,
G_2(v)  \,\, = \, \,\, \, G_2(v^2) \, \,  \, +v.
\end{eqnarray}
Modulo $\, 4$
the series $\, G_4$ is the lacunary series
 $\, \, 3 \, L(v)\, -3 \, -2 \, v$
\begin{eqnarray}
\label{Fvv4}
\hspace{-0.95in}&&   \quad \,  \,  \quad  \quad 
G_4(v)  \,\, = \, \, \, \, 
v \, \,\,  +3\,{v}^{2}\,\,  +3\,{v}^{4}\,\,  +3\,{v}^{8}\,\, 
 +3\,{v}^{16}\, +3\,{v}^{32}\, +3\,{v}^{64}\, +3\,{v}^{128}
\nonumber \\
\hspace{-0.95in}&&  \quad  \quad   \quad     \quad   \quad  \quad  \quad  \quad  \quad \, 
+3\,{v}^{256}\,\,  +3\,{v}^{512}\,\,  +3\,{v}^{1024}\,\, + \,\, \cdots 
\end{eqnarray}
which is a solution of the functional equation: 
\begin{eqnarray}
\label{Fvequ4}
\hspace{-0.95in}&&   \quad  \quad \quad   \quad  \quad  \quad  \quad   
G_4(v)  \,\, = \, \, \, G_4(v^2) \,\,\,  +v \,\,  +2 \, v^2.
\end{eqnarray}
Modulo $\,8$ the series $\, G_8(v)$  becomes more difficult to
recognise,
\begin{eqnarray}
\label{Fvv8}
\hspace{-0.95in}&&      \, \,   
G_8(v)  \,\, = \, \, \, 
v \, +3\,{v}^{2}\,+3\,{v}^{4}\,+4\,{v}^{5}\,+4\,{v}^{6}
+3\,{v}^{8}+4\,{v}^{11}+4\,{v}^{13}+4\,{v}^{14}+4\,{v}^{15}+7\,{v}^{16}
\nonumber  \\
\hspace{-0.95in}&&   \quad   \quad  
\,+4\,{v}^{17}\,+4\,{v}^{19}\,+4\,{v}^{20}\,+4\,{v}^{22}\,
+4\,{v}^{23}\,+4\,{v}^{24}\, +4\,{v}^{26}\,+4\,{v}^{27}
\,\,\, + \, \,\cdots 
\end{eqnarray}
though if we define 
\begin{eqnarray}
\label{tildeG81}
\hspace{-0.95in}&&  \quad   \, \quad \quad  \quad \quad  \quad   \quad   \quad 
{\hat G}_8(v)\, \,=\,\,\, \, G_8(v)\,\,  +L(v)\,\,  -1,
\end{eqnarray}
then
\begin{eqnarray}
\label{tildeG8}
\hspace{-0.95in}&&   \quad  \quad  \quad  \quad \quad  \quad   \quad   \quad 
 2 \cdot \, {\hat G}_8(v) \, =  \, \, \, 
 4 \, v \,\,\, \, \quad  \quad  { mod.} \,\, 8.
\end{eqnarray}
Comparing the  series  (\ref{introduce}) with the 
series $\, (\chi_H^{(1)}\, -1)/4$ which is equal to 
\begin{eqnarray}
\label{tildechi1appcompare}
\hspace{-0.95in}&&   \, \,   
\,  {{1} \over {4}} \cdot \, 
\Bigl(\Bigl({{1\, -16 \,v^4} \over {(1\, -2 \, v)^8}}\Bigr)^{1/4} \, -1\Bigr)
  \, = \, \, \,\,      
v \,\, \,  +3\,{v}^{2} \, +8\,{v}^{3}  +19\,{v}^{4} 
+44\,{v}^{5}  +100\,{v}^{6} \, +224\,{v}^{7}
\nonumber  \\
\hspace{-0.95in}&&   \quad   \quad   \, \,   
+490\,{v}^{8}  +1064\,{v}^{9} +2296\,{v}^{10}  +4928\,{v}^{11} 
 +10488\,{v}^{12}  +22180\,{v}^{13} \,\,     +  \,  \, \cdots 
\end{eqnarray}
one gets mod. $\, 2, \, 4, \, 8, \, 16, \, 32$ respectively: 
\begin{eqnarray}
\label{tildechi1app}
\hspace{-0.95in}&&   \quad 
v+{v}^{2}+{v}^{4} \quad \quad mod. \, \, 2, 
\qquad \quad \quad \quad \quad    v\,+3\,{v}^{2}+3\,{v}^{4}+2\,{v}^{8}
 \quad \quad mod. \, \, 4,
\nonumber  \\
\hspace{-0.95in}&&   \quad  
v+3\,{v}^{2}+3\,{v}^{4}+4\,{v}^{5}+4\,{v}^{6}+2\,{v}^{8}
 \quad \quad  mod. \, \, 8, 
   \\
\hspace{-0.95in}&&   \quad  
v+3\,{v}^{2}+8\,{v}^{3}+3\,{v}^{4}+12\,{v}^{5}+4\,{v}^{6}
+10\,{v}^{8}+8\,{v}^{9}+8\,{v}^{10}+8\,{v}^{12}+8\,{v}^{16}
  \quad  mod. \, \, 16.
\nonumber 
\end{eqnarray}

We see that, in contrast to the low-temperature expansion, 
 a simple rational function like $\, \chi_H^{(1)}$ yielding
polynomial expressions modulo $\, 2, \, 4, \, 8, \, 16, \, 32$
cannot give rise to the emergence of lacunary series 
like (\ref{Fvv}) and (\ref{Fvv4}). For high-temperature series, 
one must therefore rather ask whether, modulo $\, 2^r$, 
one can distinguish between the full
susceptibility $\, \chi_H$  and $\, \chi^{(1)} \, +\chi^{(3)}$.

\vskip .1cm 

\subsection{Functional equations mod. $\, 2^r$ 
for $\, \tilde{\chi}$ in the variable $\, w$.}
\label{avallowhightemp}

For the low and high-temperature series for $\, \tilde{\chi}$
in the variable $\, w$ 
(see (\ref{tildechilow}), (\ref{tildechi}), 
(\ref{tildechilow2}), (\ref{tildechi1}),  ...),
we have obtained similar results and functional equations 
modulo $\, q \, = \,2, \, 4, \, 8, \, 16, \, 32, \, 64$.
These series, and the corresponding functional equations, 
are given in \ref{appendix}. 

\subsubsection{High temperature 
for $\, \tilde{\chi}$ in the variable $\, w$. \\}
\label{whightempinw}

Let us consider the previous question of comparing 
$\, \chi_H$  with $\, \chi^{(1)} \, +\chi^{(3)}$, but
in the variable $\, w$, so that we are comparing $\, \tilde{\chi}_H$ and 
$\, \tilde{\chi}^{(1)} \, + \, \tilde{\chi}^{(3)}$ modulo $\, 2^r$.

The series expansion of the difference 
 $\, \Delta_H \, = \, \, $
$\tilde{\chi}_H \, - \,  (\tilde{\chi}_H^{(1)} \, + \, \tilde{\chi}_H^{(3)})$ 
is given in (\ref{tildechi3diff}). The series expansion
for $\, \tilde{\chi}^{(1)} \, + \, \tilde{\chi}^{(3)}$ 
can be obtained with an arbitrary number of exact coefficients,
 while $\, 2043 $ coefficients 
of the series expansion of $\, \tilde{\chi}_H$ are known.  
Considering, modulo various integers, the $\, 2043$ 
coefficients of the series (\ref{tildechi3diff}),
we found that $ \, \Delta_H\, = \, \, 0$ modulo $\, 2^r$, for $r \le 5$.

Modulo $\, 16$   one cannot distinguish $\, \tilde{\chi}_H$ 
and $\, \tilde{\chi}_H^{(1)} \, + \, \tilde{\chi}_H^{(3)}$, their series 
expansions being a very simple lacunary series:
\begin{eqnarray}
\label{tildechi3diffmodintegers16}
\hspace{-0.95in}&&      \, \, \, 
\tilde{\chi}_H(w)  
 \, \,\, = \, \, \,\,
 \tilde{\chi}_H^{(1)} \, + \, \tilde{\chi}_H^{(3)} 
 \,\, = \, \, \, \, \, 10\,w \, +8\,{w}^{3}\, +8\,{w}^{5}
 \, \,+  \,  \, 8 \, w \cdot \, L(w)
 \\
\hspace{-0.95in}&&   \quad  \,  \,\,   
 = \, \, \, 
2\,w \,+8\,{w}^{2} \,+8\,{w}^{9} \,+8\,{w}^{17} \,+8\,{w}^{33} \,
+8\,{w}^{65} 
+8\,{w}^{129} \,+8\,{w}^{257} \,+8\,{w}^{513}
 \,  \,  +\, \, \cdots
 \nonumber  
\end{eqnarray}
yielding the simple functional equation modulo 16:
\begin{eqnarray}
\label{tildechi3func16}
\hspace{-0.95in}&&  \,\,\,      \quad     \quad   \quad   \quad     \quad  
\tilde{\chi}_H(w^2) \,\, = \,\,\,\,\,
 w \cdot \, \tilde{\chi}_H(w) \, \,\,\, \,  + \, \, 8 \, w^3 \cdot \, (w^7\, -w\,-1). 
\end{eqnarray}

\vskip .1cm 

Modulo $\, 32$, similarly, one cannot distinguish $\, \tilde{\chi}_H$ 
and $\, \tilde{\chi}_H^{(1)} \, + \, \tilde{\chi}_H^{(3)}$, their series 
expansions being a  very simple lacunary series
\begin{eqnarray}
\label{tildechi3diffmodintegers}
\hspace{-0.95in}&&  \,     
2\,w \, +8\,{w}^{2} \, +8\,{w}^{9} 
\,+24\,{w}^{17} \, +24\,{w}^{33} \, +24\,{w}^{65}  +24\,{w}^{129} 
+24\,{w}^{257} +24\,{w}^{513} \,\,  +\, \, \cdots 
\nonumber  \\
\hspace{-0.95in}&&   \quad   \quad    \quad    
 \,\, = \, \, \, \, \, 
10\,w \, +16\,{w}^{2} \,+8\,{w}^{3} \,+8\,{w}^{5} \,+16\,{w}^{9}
 \, \, \,+  \,  \, 24 \, w \cdot \, L(w), 
\end{eqnarray}
where $\, L(w)$ is the lacunary series (\ref{lacunary}).
This yields the simple functional equation modulo 32:
\begin{eqnarray}
\label{tildechi3func16}
\hspace{-0.95in}&&  \,\,\,      \quad     \quad     \quad     \quad  
\tilde{\chi}_H(w^2) \,\,\, = \,\,\,\,\,
 w \cdot \, \tilde{\chi}_H(w)
 \, \,\, \,  + \, \,  8 \, w^3 \cdot \, (2\,w^{15} \, -w^7 \,+w \,-5 ). 
\end{eqnarray}

Modulo $\, 64$,  $\, 128$, similarly, one cannot distinguish between $\, \tilde{\chi}_H$ 
and $\, \tilde{\chi}_H^{(1)} \, + \, \tilde{\chi}_H^{(3)}\, + \, \tilde{\chi}_H^{(5)}$, 
their series expansions being, again, very simple lacunary series.

\vskip .1cm 

\subsubsection{Low temperature 
for $\, \tilde{\chi}$ in the variable $\, w$. \\}
\label{wlowtempinw}

Similarly, if one compares the low-temperature full susceptibility with
$\, \tilde{\chi}_L^{(2)}(w)$ modulo $\, 32$ one finds the lacunary series:
\begin{eqnarray}
\label{tildechi3func16}
\hspace{-0.95in}&&  \,\,\,   \quad  \quad    
\, \tilde{\chi}_L(w) \,\, =  \,\,\,\,  
4\,{w}^{4}\,\, +16\,{w}^{6}\,+24\,{w}^{8}\,+16\,{w}^{12}\,+16\,{w}^{20}\,
+16\,{w}^{36}\,+16\,{w}^{68}\,
\nonumber \\
\hspace{-0.95in}&&   \quad  \quad  \quad  \quad  \quad \quad    \quad   \quad 
\,+16\,{w}^{132}\, +16\,{w}^{260} \, +16\,{w}^{516} \, +16\,{w}^{1028} 
\, \, \, + \,\,\cdots,  
\end{eqnarray}
versus 
\begin{eqnarray}
\label{tildechi3func16}
\hspace{-0.95in}&&  \,\,\,      \quad   
\, \tilde{\chi}_L^{(2)}(w)\,\, =  \,\,\,\,  
4\,{w}^{4}\,  \,  +16\,{w}^{6}\,  +24\,{w}^{8}\,  +16\,{w}^{12}\,  +16\,{w}^{16}\, 
 +16\,{w}^{20}\,  +16\,{w}^{36}\,  +16\,{w}^{68}\,
\nonumber \\
\hspace{-0.95in}&&   \quad  \quad \quad  \quad   \quad  \quad   \quad 
    +16\,{w}^{132}\, +16\,{w}^{260} \, +16\,{w}^{516} \, +16\,{w}^{1028} 
\,\, \, \,  + \,\,\cdots,   
\end{eqnarray}
the difference being only $\, 16\, w^{16}$. 

One finds that  the difference 
between $\, \tilde{\chi}_L$ and  $\, \tilde{\chi}_L^{(2)}$, given in eqn. (\ref{tildechi2diff}),
is zero modulo $\, 2, \, 4, \, 8, \, 16$,  and equal to
$\, 16\, w^{16}$ modulo $\, 32$. Modulo $\, 64$ it is given 
by a lacunary series 
\begin{eqnarray}
\label{becomes}
\hspace{-0.95in}&& \,  \quad  \quad  \quad  \quad 
 \tilde{\chi}_L \, - \, \tilde{\chi}_L^{(2)}
 \,  \,  \, =  \,\, \, \,  \,32\, w^4 \cdot L(w) \, \,  \, +16\, w^{16} \,  \,
 \nonumber  \\
\hspace{-0.95in}&& \quad \quad  \quad  \quad \quad  \quad  \quad  \quad  \quad 
+ \,  32\, w^4 \cdot \, (w^{28} \, -w^8 \, -w^4 \, -w^2 \, -w \, -1). 
\end{eqnarray}
One finds that the difference between
 $\, \tilde{\chi}_L$ and  
$\, \tilde{\chi}_L^{(2)}\, + \,\tilde{\chi}_L^{(4)}$, 
as given by (\ref{tildechi24diff}), is zero
modulo $\, 2, \, 4, \, 8, \, 16, \, 32, \, 64$ and 
is given by a lacunary series modulo $\,128$:
\begin{eqnarray}
\label{becomes2}
\hspace{-0.95in}&& \quad \quad  \quad   \quad 
 \tilde{\chi}_L \, - \, (\tilde{\chi}_L^{(2)}\, + \, \tilde{\chi}_L^{(4)}) 
 \,  \,  \, =  \,\, \, \, 64 \, w^4 \cdot L(w) \, \,  \,
 \nonumber  \\
\hspace{-0.95in}&& \quad \quad \quad  \quad  \quad   \quad   \quad  \quad   \quad 
- \,  64 \, w^4 \cdot \, (w^{16} \, +w^8 \,+w^4 \,+w^2 \,+w \, +1). 
\end{eqnarray}
If one includes $\,\tilde{\chi}_L^{(6)}$,  the difference 
(\ref{tildechi246diff})
between $\, \tilde{\chi}_L$ and  
$\, \tilde{\chi}_L^{(2)} \, + \, \tilde{\chi}_L^{(4)}\, + \, \tilde{\chi}_L^{(6)}$
is seen to be zero modulo 
$\, 2, \, 4, \, 8, \, 16, \, 32, \, 64, \, 128, \, 256$.

The scenario seems to be that one cannot distinguish 
between the series for  $\, \tilde{\chi}_L$  and that for a 
{\em finite sum} like  
 $\, \tilde{\chi}_L^{(2)}\, + \,\tilde{\chi}_L^{(4)}$
$ \, + \, \cdots \, +\, \tilde{\chi}_L^{(2n)}$
modulo $\, 2^r$ where $\, r$ grows 
with $\, n$. The series expansion for the $\, \tilde{\chi}_L^{(n)}$  are given 
in~\cite{Iwan2}, up to $\, n\, = \, \, 12$. This scenario has been 
 checked and found to hold up to $\,\tilde{\chi}_L^{(12)}$. Recall that the {\em finite sum} 
 $\, \tilde{\chi}_L^{(2)}\, + \,\tilde{\chi}_L^{(4)}$
$ \, + \, \cdots \, +\, \tilde{\chi}_L^{(2n)}$
is {\em also the diagonal of a rational function}~\cite{Short}, implying that
this finite sum {\em reduces to algebraic functions modulo primes, 
or power  of primes}, and thus {\em satisfies functional equations modulo primes, 
or power  of primes}. For instance, $\, \tilde{\chi}_L$, which cannot
be distinguished from this sum modulo $\, 2^r$ for some $r$, 
satisfies  a functional equation modulo $\, 2^r$. 

These functional equations can thus be seen as 
related to the functional equations for $\, \tilde{\chi}_L^{(n)}$.
For instance modulo $\, 2$, $\, \tilde{\chi}_H^{(3)}(w)/8$ becomes 
the lacunary series 
\begin{eqnarray}
\label{becomes3}
\hspace{-0.95in}&& \quad \quad  \quad  \quad  \quad  \quad 
{{ \tilde{\chi}_H^{(3)}}(w) \over {8}} 
\,\,\, = \, \, \, \, \,  
w \cdot \, L(w) \,\,\, \,  -w\cdot \, (w^4 \, +w^2 \, +w \, +1),  
\end{eqnarray}
from which one deduces the functional equation modulo $\, 2$: 
\begin{eqnarray}
\label{deduces3}
\hspace{-0.95in}&& \quad \quad  \quad  \quad  \quad 
w \cdot \, {{ \tilde{\chi}_H^{(3)}(w)} \over {8}} 
 \,\,  \,\, = \, \, \, \, \,\,  {{ \tilde{\chi}_H^{(3)}(w^2)} \over {8}} 
\,\, \,\, + \, \, \, \,  w^{10}.  
\end{eqnarray}
Modulo $\, 4$ the series
$\, \tilde{\chi}_H^{(3)}(w)/8$ becomes the lacunary series 
\begin{eqnarray}
\label{becomes4}
\hspace{-0.95in}&& \quad \quad  \quad  \quad  \quad 
{{ \tilde{\chi}_H^{(3)}(w)} \over {8}} 
\,\, = \, \,\, \, \, 3 \,  w \cdot \, L(w) \,  \,\,\, 
+ w \cdot \, (2\, w^8 \, + \, w^4 \, +w^2 \, +w \, +1),  
\end{eqnarray}
from which one deduces the functional equation modulo $\, 4$: 
\begin{eqnarray}
\label{deduces4}
\hspace{-0.95in}&& \quad \quad  \quad  \quad  \quad 
w \cdot \, {{ \tilde{\chi}_H^{(3)}(w)} \over {8}} 
 \,  \,\, = \, \, \,\, \, \, 
 {{ \tilde{\chi}_H^{(3)}(w^2)} \over {8}} 
\,\,\,  + \, \, \,  w^3 \cdot \, (4 \, +w^7 \, -2\, w^{15}).  
\end{eqnarray}

For  $\, \tilde{\chi}_L^{(4)}/16$ we have similar results. 
The series $\, \tilde{\chi}_L^{(4)}/16$ reduces, modulo $\, 2$,
 to $\, w^{16}$.  Modulo $\, 4$ the series
$\, \tilde{\chi}_L^{(4)}/16$ becomes\footnote[1]{Here,
the calculations can be checked with an arbitrary number of coefficients. 
We did so with $\, 6000$ coefficients. }  the 
simple lacunary series 
\begin{eqnarray}
\label{becomes5}
\hspace{-0.95in}&& \, 
 {{ \tilde{\chi}_L^{(4)}(w)} \over {16}} 
 \,  \,\, = \, \, \, \, \,  2 \,  w^4 \cdot \, L(w) 
\,   \, \, +w^{16} \,   \,  
+ 2 \, w^4 \cdot \, (w^{28} \, +\, w^8 \, + \, w^4 \, +w^2 \, +w \, +1), 
\end{eqnarray}
yielding the functional equation  modulo $\, 4$ 
\begin{eqnarray}
\label{deduces5}
\hspace{-0.95in}&& \, \quad \,  \,\,
 {{ \tilde{\chi}_L^{(4)}(w^2)} \over {16}} 
 \,  \,\, = \, \, \, \, \,  
w^4 \cdot \,  {{ \tilde{\chi}_L^{(4)}(w)} \over {16}} \,\, \, 
+  \, w^{20} \cdot \, (2\, w^{44} \, -\,2\, w^{16} + w^{12} \, + \,2\, w^4 \, -1).
\end{eqnarray}

\vskip .1cm 

\section{Automaton interpretation of the functional equations}
\label{automaton.}

 Recalling the decomposition
of the full susceptibility as an infinite sum of $\,n$-fold integrals,  
$\, \chi^{(n)}$, these striking results can be seen as a consequence of 
the fact that, {\em modulo integers that are powers of the prime $\, 2$}, 
the full susceptibility series is the same lacunary series as the 
series for the first $\, \chi^{(n)}$'s: for instance the low-temperature 
series modulo $64$ of the  full susceptibility  series and  of 
$\, \tilde{\chi}^{(2)} \, +\tilde{\chi}^{(4)}$ 
(which is the diagonal of a rational function) are the same. There is 
a not-widely-known  {\em discrete automaton}~\cite{Poorten,Rowland} 
result that, modulo a prime $\,p$, 
{\em diagonals of  rational functions}~\cite{Christol} 
not only {\em reduce to algebraic functions}, 
but also satisfy~\cite{Poorten}  ``functional equations modulo $\, p^r$'' 
of the form $\, F(f(x), \,f(x^p), \, \cdots,\,  f(x^{p^h}))\,= \,\, 0$. 

\vskip .1cm 

Let us recall some relevant results on discrete 
automaton~\cite{Poorten,Rauzy,Denef,Adamczewski}:
 modulo a prime $\, p$,
 the diagonal of a rational function reduces to an algebraic 
function, and this is also true modulo $\, p^r$ ($\, p$ prime, 
$\, r$ integer). Furthermore, these papers
tell us that if $\,f(x)$ is 
algebraic modulo a prime $\, p$  then 
$\, 1, \, f(x), \, f^2(x), \, f^3(x), \cdots $ are linearly dependent,
and $\, 1, \, f(x), \, f^p(x), \, f^{p^2}(x),  \, f^{p^3}(x), \cdots $
are also linearly dependent. From Fermat's little theorem,
namely that if $\, p$ is a prime number, $ \,a^p \, = \, a$, (mod. $p$),
one deduces for any series $\, f(x) \, = \, \sum a_n \cdot \, x^{n}$
\begin{eqnarray}
\label{heavily}
\hspace{-0.95in}&&    \quad   \, \, \quad \quad  
(\sum a_n \cdot \, x^n)^p \, \, = \, \, \,
 \sum a_n^p \cdot \, x^{p \,n}\, \, = \, \, \,
\sum a_n \cdot \, x^{p \,n},
 \quad \quad \quad 
mod. \, \,  p, 
\end{eqnarray}
and, thus, $\, f(x)^p   \, \, = \, \, \,    f(x^p) \, $ modulo  $\, p$, and, 
more generally, $\,f(x)^{p^r}  \, \, = \, \, \,  f(x^{p^r})\, $ 
modulo  $\, p^r$. One deduces that the relations
$\,\, F(f(x), \,f(x^p), \, \cdots,\,  f(x^{p^h}))\,= \,\, 0 \, $
can, in fact be written {\em linearly}, as
\begin{eqnarray}
\label{linearly}
\hspace{-0.95in}&&   \quad  \quad \quad  \quad  \quad  \quad    \quad  \quad 
\sum_n \, p_n(x) \cdot \, f(x^{p^n})
 \,\,= \,\,\, 0, 
\end{eqnarray}
where the $\, p_n(x)$ are polynomials with integer coefficients, 
(see for instance section 2 in Lipshitz and van der Poorten~\cite{Poorten}). 

\vskip .1cm 

Series generated by a {\em finite automaton} correspond to a 
{\em system of algebraic equations}, which correspond, in turn 
(non trivially) to {\em being algebraic}. All these 
{\em functional equations occurring for discrete automata} can be seen as  
{\em functional equations associated with algebraic functions modulo integers}, 
in particular {\em diagonals of rational functions}. This can be seen as 
the origin of the functional equations of this paper. The 
functional equations we have obtained can be 
interpreted\footnote[2]{Equivalently, our 
conjectured functional equations can be seen as 
conjectures on the fact that, for instance, 
the {\em non-holonomic infinite sum} 
$\, \tilde{\chi}_L \, -(\tilde{\chi}_L^{(2)} +\tilde{\chi}_L^{(4)}) \,\, $
$=\, \,\tilde{\chi}_L^{(6)} \, + \tilde{\chi}_L^{(8)} \, + \,\,\cdots $
reduces to zero  modulo $64$, and possibly, that each series 
$\,\tilde{\chi}_L^{(2\, n)}$ for $\, n\, \ge 3$, 
reduces to zero modulo $64$, which corresponds to the 
(experimental) remark of section (\ref{definition}),
that the  $\, \tilde{\chi}_L^{(n)}(w)$ 
(resp. $\, \tilde{\chi}_H^{(n)}(w)$)
have an overall factor $\, 2^{n}$.} 
as consequences of the fact that, 
{\em modulo some integers that are powers of the prime $\, 2$}, one 
{\em cannot really make a distinction between the full susceptibility
 and the diagonal of a rational function} (like the sum of the 
first $\, \chi^{(n)}$s),
and consequently reduce to algebraic functions modulo $\, 2^r$. One could 
say that {\em non-holonomic functions}, like the full susceptibility
of the Ising model, correspond to {\em ``almost diagonal functions''}.

\vskip .1cm 

The automaton interpretation of this section can be revisited from a 
{\em binomial} viewpoint. Recall that the coefficients of the 
series expansion of diagonals of rational functions 
{\em necessarily reduce to nested sums of products of binomials}~\cite{Lairez,Lairez1}. 
{\em Binomial coefficients modulo prime powers} have been considered 
by many great mathematicians of the nineteenth century\footnote[8]{For instance
 Cauchy, Cayley, Gauss, Hensel, Hermite, Kummer, 
Legendre, see~\cite{Dickson,Granville}. The 
study of {\em congruences of combinatorial numbers}~\cite{Sun} usually starts with 
their {\em $\, p$-adic order}: it was first studied by Kummer~\cite{Kummer}.},
yielding a large set of elegant results.
Among the various prime powers, the powers of $\, 2$ seem to play a 
selected\footnote[9]{In 1899 Glaisher observed that the number of odd entries
in any given row of Pascal's triangle is a power of $\, 2$.}
role~\cite{Kauers}. Combining these two set of results is another
approach to the main problem addressed in this paper, namely
the study of (infinite sums of) diagonals of rational functions 
modulo prime powers.

For powers of the prime $\, 2$, the functional equations satisfied 
by the full susceptibility are quite simple ones, which are associated with 
the lacunary series\footnote[1]{Note that, as far as reduction to algebraic 
functions modulo  powers of the prime $\, 2$ is concerned, a remarkably simple
quadratic algebraic function corresponding to the Catalan number generating 
function also reduces to this lacunary series (\ref{lacunary}), 
as can be seen in \ref{heuristic}.} (\ref{lacunary}). 
Of course for powers of other primes ($3^r$, $\, 5^r$, ...), the functional 
equations satisfied by the full susceptibility should, if they exist,
be much more involved, certainly not reducing to simple lacunary series.
For powers of other primes, the scenario that
modulo some powers of primes,  
one cannot disentangle the full susceptibility from 
some finite sum of $\, \chi^{(n)}$'s, is {\em no longer valid}. For instance,
if one considers the series expansion (\ref{tildechi5diff})
of the difference $\tilde{\chi}_H \, $
$- \,  (\tilde{\chi}_H^{(1)} \, + \, \tilde{\chi}_H^{(3)})$, one sees that
the coefficient of $\, w^{100}$ and $\, w^{101}$ are, respectively, the following 
products of primes\footnote[5]{Using the command ``ifactor'' in Maple.}: 
\begin{eqnarray}
\label{ifactor}
\hspace{-0.95in}&&   \, \,   \quad   
2^{12} \cdot \, 59 \cdot \, 1403746269427 \cdot \,1965616530023 
   \cdot \,   269691689798741092891, 
\nonumber \\
\hspace{-0.95in}&&   \, \,   \quad   
2^{8} \cdot \, 29 \cdot \, 26811049 \cdot \, 
99658008281797903856656009433736710068597. 
\end{eqnarray}
Similarly, if one considers the series expansion (\ref{tildechi3}) of 
$\, \tilde{\chi}_H^{(3)}/8$, one finds that the coefficient of 
$\, w^{100}$ and $\, w^{101}$ are, respectively, the following 
products of primes: 
\begin{eqnarray}
\label{ifactor2}
\hspace{-0.95in}&&   \, \,   \quad   
2^2 \cdot \, 3^2 \cdot \, 263 \cdot \, 3291604173673 \cdot \, 340864762033 \cdot \, 3935416959987419344918432619, 
\nonumber \\
\hspace{-0.95in}&&   \, \,   \quad   
2^3 \cdot \, 5 \cdot \, 
5581 \cdot \, 1400518348065785091954773485695960563962083118761649. 
\end{eqnarray}
Besides powers of $\, 2$, there is no prime cancelling all the coefficients 
of the difference (\ref{tildechi5diff}) or of $\,\tilde{\chi}_H^{(3)}$. 

\vskip .1cm 

\vskip .1cm 

The question whether modulo primes different from $\, 2$,
or powers of primes {\em different from} $\, 2^r$, 
the full susceptibility, that no longer reduces to the sum of the first
$\, \tilde{\chi}^{(n)}$'s,  satisfies
(probably involved) functional equations remains open.

\vskip .1cm 

\section{Comments and speculations.}
\label{specul}

\subsection{Towards a physical interpretation of the functional equations.}
\label{physical}

We can view these exact functional equations modulo
integers that are powers of the prime $\, 2$, as a 
{\em finite discrete  automata}~\cite{Rowland} result corresponding 
to the fact that, modulo such integers,
one cannot disentangle the full susceptibility from the diagonal 
of a rational function. From a more speculative, but more physical 
perspective, one might hope that 
such functional equations are the ``shadow'' modulo 
primes or powers of primes, of (probably very involved) 
functional equations\footnote[2]{In characteristic zero, 
not ``modulo primes or powers of primes''.}, 
 the $\, x \, \rightarrow \, x^p \,\, $ Frobenius symmetry being, 
in fact, an {\em infinite order transformation}. Furthermore, 
such an {\em infinite order discrete transformation} 
might be seen as a {\em symmetry} of the model. Along this symmetry
line,  recall that, in the case of the square 
Ising model, any {\em isogeny} of the elliptic curve
parametrizing the model~\cite{CalabiYauIsing} can be interpreted as 
an {\em exact generator of the renormalization group}~\cite{Renorm}.  

We remark that $\, \tilde{\chi}_L^{(2)}$  (see (\ref{tildechilow2})),
can be written so that the Landen modulus clearly appears.
Consider the $\, _2F_1$ hypergeometric function $\,\Phi(x)$, 
and recall the  Landen modulus $\, k_L$:
\begin{eqnarray}
\label{Phi}
\hspace{-0.95in}&&   \, \,   \quad    \quad  \quad    \quad  
\Phi(x)  \,\, = \, \,  \,  \, \,
{{1} \over {4^3}} \, 
 \cdot \, x^4  \cdot \,
 _2F_1\Bigl(([{{3} \over{2}}, \,{{5} \over{2}}],[3], \, x^2 \Bigr), 
\qquad \quad   \quad  
 k_L \, = \, \, {{ 2 \, \sqrt{k}} \over {1\, +k }}.
\nonumber 
\end{eqnarray}
From  (\ref{tildechilow2}) $\, \tilde{\chi}_L^{(2)}$  reads: 
\begin{eqnarray}
\label{chilow2ter}
\hspace{-0.95in}&&     \quad  \, 
 \tilde{\chi}_L^{(2)} \,\, = \, \, \,  \Phi(k_L) \,\,\,   = \, \, \,  \Phi(4 \, w) 
\,\, = \, \, \,  \Phi\Bigl({{ 4 \, v} \over {1\, +4\, v^2 }}\Bigr) 
\,\, = \, \, \,  \Phi\Bigl({{ 2 \, s} \over { 1\, + s^2}}\Bigr)
 \\
\hspace{-0.95in}&&   \quad \,  \quad   
 = \, \, \,
4\,{v}^{4} \, +16\,{v}^{6} \, +120\,{v}^{8} \,
 +480\,{v}^{10} \, +2800\,{v}^{12} \, 
+11200\,{v}^{14} \, +58800\,{v}^{16}
\,\,\, + \, \, \cdots 
\nonumber 
\end{eqnarray}
One would like to see the 
{\em Landen transformation} $\, k \, \rightarrow \, k_L$, 
which can be viewed as an {\em  exact generator}\footnote[1]{This highly
selected infinite order 
transformation (isogeny of the elliptic curve parametrizing
the model~\cite{CalabiYauIsing,Renorm}) has $\, k\, = \, \, 1$ as a fixed point, 
$\, k\, = \, \, 0, \, \infty$ being clearly special~\cite{Renorm}. }
of the  {\em renormalization group}~\cite{Renorm},  {\em as a symmetry of} 
$\, \tilde{\chi}_L^{(2)}$,  
{\em before seeing it as a symmetry of the full susceptibility}. Remarkably, 
this is the case:
$\, \Phi(k_L)$ and $\, \Phi(k)$ are very closely and very simply related!
This remarkable relation can be written in many ways, using the various 
variables we have introduced~\cite{Khi6,CalabiYauIsing,High,bernie2010}
 ($s$, $\, k$, $\, w$, $\, v$), but since 
our functional relations are mostly written as series in $\, v$,
we will write this  relation in $\, v$. In $\, v$ the {\em Landen transformation} 
corresponds to 
\begin{eqnarray}
\label{Landenv}
\hspace{-0.95in}&&   \quad   \quad     \quad   \quad  \quad 
 4 \, v^2 \, \, \,\,  \longrightarrow \, \, \, \, \, 
 {{4 \, v} \over {1 \, + \, 4 \, v^2}} 
\qquad \hbox{or:} \qquad  \quad  
 v^2 \, \, \,\,  \longrightarrow \, \, \, \, \,
  {{ v} \over {1 \, + \, 4 \, v^2}}.
\end{eqnarray}
Let us introduce 
\begin{eqnarray}
\label{Phi}
\hspace{-0.95in}&&   \quad  \quad   \quad \quad 
 \quad    \quad   \quad    \quad     \quad  
\Psi(x)  \,\, = \, \,  \,  \, \, 
 {{1 \, + \, x} \over {x}} \cdot {{d \Phi(x)} \over {d x}}.
\end{eqnarray}
One then has the remarkably simple (differential-functional equation) 
relation representing the {\em Landen transformation as a 
symmetry of} $\, \tilde{\chi}_L^{(2)}$:
\begin{eqnarray}
\label{Phi}
\hspace{-0.95in}&&   \quad \quad    \quad \quad  \quad
    \quad  \quad    \quad   \quad    
\Phi\Bigl( {{4 \, v} \over {1 \, + \, 4 \, v^2}}  \Bigr) 
   \,\, = \,\, \,  \,  \, \,  4 \cdot \,  \Psi(4 \, v^2). 
\end{eqnarray}

\vskip .1cm 

Pursuing this line of argument on functional equations
 with an {\em infinite order} transformation
(hopefully with a physical symmetry interpretation like the
{\em Landen transformation 
representing a generator of the renormalization group}~\cite{Renorm}), 
it is tempting to 
imagine the $\, v\, \rightarrow \,  v^2$ infinite order
transformation in functional equations like (\ref{goodchilowmod32}),
(\ref{Fvequ}), or (\ref{Fvv4}), as a mod. $\,2^r$ reduction of an 
{\em infinite order symmetry} of the model. In such a scenario, since one 
cannot distinguish, modulo $\, 2$  or $\, 4$, between $\, v$ 
and $\, v/(1\, + \, 4 \, v^2)$, a functional equation
$\, G_2(v)  \,\, = \, \,\, \, G_2(v^2) \, +v$, 
like (\ref{Fvequ}), could also be written as 
\begin{eqnarray}
\label{could}
\hspace{-0.95in}&&   \quad     
 G_2\Bigl({{v} \over {1 \, + \, 4 \, v^2}} \Bigr)
   \,\, = \, \,\,\, \, G_2(v^2)   \, +v 
\quad \, \,  \,   \,    \,  \,    \hbox{or:}  \, \,      \quad \quad 
G_2\Bigl({{k_L} \over {4}} \Bigr)\, = \, \,\,\, G_2\Bigl({{k} \over {4}} \Bigr)
\, + \, \, {{\sqrt{k}} \over {2}}, 
\end{eqnarray}
and one could expect that the functional equations we discover
modulo  $\,2^r$, are also the restriction modulo $\,2^r$
of some (quite involved) functional equations where the 
infinite order transformations have some physical meaning.
Keeping in mind the unit circle 
{\em natural boundary} of the full susceptibility of the 
Ising model, it is worth recalling that functional equations like 
$\, G(v)  \,\, = \, \,\,  G(v^2)  \, +v$, not modulo integers
but in characteristic zero, are the simplest 
examples to actually show that a series has a 
(unit circle) {\em natural boundary}. 

\vskip .1cm 

Recalling the expression of $\tilde{\chi}_L^{(2)}(w)$
given by (\ref{tildechilow2}), the previous  
functional equation (\ref{Phi}) reads:
\begin{eqnarray}
\label{Phibis}
\hspace{-0.95in}&&   
\quad \quad     \quad   \quad  \quad  \quad    
\tilde{\chi}_L^{(2)}\Bigl({{v} \over {1 \, +4 \,v^2}} \Bigr) 
\, \,\, = \,\,\,\, \, {{ 1 } \over { 8}}\, 
\cdot \, 
 {{1 \, +4 \,v^2} \over {v^3}} \cdot \, {{ d \tilde{\chi}_L^{(2)}(v^2)} \over { d v}}, 
\end{eqnarray}
This might suggest replacing $\tilde{\chi}_L^{(2)}(w)$
by $\,  \tilde{\chi}_L(w)$ so as
 to consider this differential-functional equation (\ref{Phibis}) 
for the full susceptibility 
given by (\ref{tildechilow}) modulo $\, 2, \, 4,  \, 8$.
Unfortunatly, in contrast with  the calculations performed 
in section (\ref{avallowtemp}), one finds that a 
{\em differential}\footnote[2]{Therefore
different from the functional equations in section (\ref{avallowtemp}).}
 functional equation like (\ref{Phibis}) is not  
satisfied by the full susceptibility modulo $\, 2, \, 4, \, 8$.
This seems to suggest that, if a ``master'' functional equation 
with an infinite order transformation symmetry exists (in exact 
arithmetic, not modulo some integers) for the full 
susceptibility, it is certainly much more involved that any 
simple generalization of (\ref{Phibis}).

\vskip .1cm 

Of course, all these ideas are quite speculative. It is reasonable to imagine that
there must be some (probably involved) representation of the renormalization
group~\cite{Renorm} of the full susceptibility. In particular, for this integrable
model which has an elliptic parametrization, {\em one might expect a representation of
the action of the Landen transformation on the full susceptibility}.

 \vskip .2cm 

\subsection{Non-holonomic functions that are ratios of 
diagonals of rational functions, and beyond.}
\label{beyon}

Almost everything remains to be done to understand and describe
this class of ``nice'' non-holonomic functions reducing to algebraic functions
modulo some powers of primes. It is worth recalling that,
while the product of two holonomic functions is a holonomic function, the {\em ratio of 
two holonomic functions is, in general\footnote[1]{Except when
 the holonomic function in the denominator is an algebraic function: in that case 
the ratio is also holonomic.}, non-holonomic}! The class of functions that are 
expressible as a {\em ratio of 
two holonomic functions}, and, further, {\em the ratio of diagonals of rational functions},
is clearly a {\em very interesting class of functions}: they are such that 
their series can be recast into series with {\em integer coefficients}~\cite{Short,Christol} 
(the ratio of series with integer coefficients is up to an overall integer a series 
with integer coefficients), and that  their series, modulo primes, or 
modulo powers of primes, {\em reduce to algebraic 
functions}\footnote[5]{More generally, a rational or even algebraic
 function  (with integer coefficients) of  holonomic functions,
 is such that {\em it reduces modulo primes, 
or modulo powers of primes, to an algebraic function}.} (the ratio of series 
reducing to algebraic functions reduces to  the ratio of algebraic functions, 
and thus reduces to algebraic functions). Keeping in mind  
the (natural boundary of the) susceptibility of the Ising model,
it is worth recalling that the ratio of holonomic functions
can also yield a {\em natural boundary}, as can be seen 
from the solutions of {\em non-linear} Chazy III equations~\cite{Chazy1,Chazy2}.

The solutions of a particular {\em non-linear} third order differential equations 
having the Painlev\'e property, the Chazy III 
equations~\cite{Chazy1,Chazy2}, 
have (circular) natural boundaries, and this can be 
seen as a {\em direct consequence} of the fact that 
the solutions correspond to the ratio of two holonomic 
functions, as shown by Chazy in crystal clear papers.

The {\em Chazy III equation}~\cite{Chazy1,Chazy2}
 is a third-order {\em non-linear}
 differential equation  with a movable singularity 
that has a {\em natural boundary}
 for its solutions~\cite{Chakravarty}:
\begin{eqnarray}
\label{Chazy}
\hspace{-0.85in}&& \quad \quad \quad \quad  \quad \quad \quad \quad 
{{d^3 y} \over {dx^3}} \,\,  = \, \, \, \, 
2 \, y \cdot \, {{d^2 y} \over {d x^2}} \, \,
 - 3 \, \Bigl({{d y} \over {d x}} \Bigr)^2. 
\end{eqnarray}
It can be rewritten in terms of a {\em Schwarzian derivative}:
\begin{eqnarray}
\label{Chazy2}
\hspace{-0.95in}&& \quad \quad \quad \, 
 \, f^{(4)} \, \, = \, \, \,  2 \, f'^2 \cdot \, \{f, \, x\} \, \, = \, \, \,  
2 \, f' \, f'''\, - 3 \, f''^2   \,  \quad \quad
\hbox{with:} \quad \quad  \quad
y \, = \, \, {{d f} \over {dx}}.
\end{eqnarray}
Similarly, it is important to recall
 that the {\em ratio} of {\em two holonomic} functions, which is in general
 a {\em non-holonomic} function, is the solution of a non-linear 
{\em Schwarzian derivative} ODE:
\begin{eqnarray}
\hspace{-0.95in}&&  \quad  \quad  \quad 
{{d^2 y} \over {dx}}\, +R(x) \cdot \, y \, = \, \, 0, \,  \,  \quad \quad  
\tau(x) \, = \, \, {{y_1} \over {y_2}},  \, \, \quad \quad \,\,\,
 \{\tau(x), \, x\} \, \, = \, \, \, 2 \, R(x). 
\end{eqnarray}
The Chazy III non-linear differential equation (\ref{Chazy}) 
 has the {\em quasi-modular form} Eisenstein series $\, E_2/2$. It 
can also be written 
as a log-derivative\footnote[2]{One takes the derivative with respect 
to the nome $\, q$ (see equation (6) in~\cite{Chakravarty}). }, namely 
a {\em ratio} $\, \Delta'/\Delta$, where
 {\em Ramanujan's modular discriminant function}~\cite{Ramanujan,Ramanujan2}  
$\, \Delta$ is actually a selected holonomic function: a {\em modular form}.

 \vskip .1cm  

It is worth recalling, with the example of the enumeration of 
{\em three-dimensional convex polygons}~\cite{Bousquet},  
that we have already encountered, in enumerative combinatorics,  
the emergence of {\em ratios of holonomic functions}. The class of functions characterised by ratios of 
holonomic functions and ratios of diagonals of rational functions
 is certainly an over-simplified scenario 
for the susceptibilitity of the Ising model. It is however an interesting 
``toy class'' for the susceptibilitity of the Ising model, the 
class of the {\em algebraic functions of diagonals of rational functions}
 being much too large to reasonably explore. 

 \vskip .1cm 

\section{Conclusion}
\label{conclusion}

This paper underlines the central role of {\em discrete finite automata}, or 
{\em diagonals of rational functions}, in lattice statistical 
mechanics and enumerative combinatorics, in particular regarding
the challenging problem of the full susceptibility of the 
two-dimensional Ising model~\cite{Orrick}. 

The natural emergence of {\em diagonals of rational functions}
 in an extremely large set 
of lattice statistical mechanics and enumerative combinatorics models, has been 
emphasised and explained in~\cite{Short}. That paper explains why a large
class of functions describing lattice models that 
can be expressed as $\, n$-fold integrals
of an algebraic\footnote[5]{Or even holonomic.} integrand~\cite{Short}, 
which are, consequently, solutions of linear differential equations, and, 
thus, at first sight, {\em transcendental} functions, is in fact a remarkable 
class of {\em transcendental} holonomic functions, 
namely {\em diagonals of rational functions}~\cite{Short}. 

The corresponding selected 
linear differential operators are not only Fuchsian, but 
also~\cite{bo-bo-ha-ma-we-ze-09} globally 
nilpotent\footnote[1]{Their critical exponents are rational numbers, their Wronskian are
$N$-th roots of rational functions, etc.}, and, since these transcendental functions are 
{\em diagonals of rational functions}, they reduce to 
{\em algebraic functions modulo any prime}~\cite{Short}.
They even reduce to {\em algebraic functions modulo any integral power of a prime number}. 
We may call this class of transcendental {\em holonomic} functions, that quite naturally 
occur in so many problems of theoretical physics~\cite{Short}, ``almost algebraic functions''. 

As far as {\em transcendental non-holonomic functions} are concerned, 
the full susceptibility of the two-dimensional Ising model is 
``algorithmically integrable'' (with an $\,O( N^4)$ polynomial algorithm)
and can be decomposed  as an infinite sum of $\,n$-fold integrals,
that have been shown to be {\em diagonals of rational functions}~\cite{Short}.
Such nice  {\em transcendental non-holonomic functions} emerging in physics
require further concepts and tools to characterize and analyze them. 
 
 \vskip .1cm 

In this paper we have obtained {\em exact functional equations} for 
 low and high temperature series of the full susceptibility {\em modulo integers that
are powers of the prime $\, 2$}, the series being associated with
 simple lacunary series. Since these 
exact results come from remarkably long low- and high-temperature series~\cite{Iwan} 
with more than 2000 coefficients, these exact functional equations are currently
not yet proved but extremely plausible conjectures. Recalling the decomposition
of the full susceptibility as an infinite sum of $\,n$-fold integrals $\, \chi^{(n)}$,
these striking results can, in fact, be seen as a consequence of the fact that, 
{\em modulo integers that are powers of the prime $\, 2$}, the full susceptibility  
series are the same series as the series for the sum 
of the first $\, \chi^{(n)}$'s: for instance 
the low-temperature series modulo $16$ of the  full susceptibility  series and  
of $\, \chi^{(2)}$ (which is the diagonal of a rational function) are the same. 

Modulo a prime $\,p$, 
{\em diagonals of a rational function} not 
only reduce to algebraic functions, but also satisfy  
equations of the form $\, F(f(x), \,f(x^p), \, \cdots,\,  f(x^{p^h}))\,= \,\, 0$. In 
other words, the functional equations we have obtained, can be interpreted
as the fact that 
{\em modulo some integers that are powers of the prime $\, 2$}, one 
{\em cannot really distinguish between the full susceptibility
 and the diagonal of a rational function} 
(like, for instance, $\, \chi^{(2)}+\chi^{(4)}$, ...). The scenario seems to be 
that one cannot distinguish the series for  $\, \tilde{\chi}_L$ 
and for a finite sum like  
 $\, \tilde{\chi}_L^{(2)}\, + \,\tilde{\chi}_L^{(4)}$
$ \, + \, \cdots \, +\, \tilde{\chi}_L^{(2n)}$
modulo $\, 2^r$ where $\, r$ grows 
with $\, n$. The series expansion for the $\, \tilde{\chi}_L^{(n)}$  are given 
in~\cite{Iwan2}, up to $\, n\, = \, \, 12$. This scenario can 
be checked up to $\, \tilde{\chi}_L^{(12)}$. Recall that the finite sum 
 $\, \tilde{\chi}_L^{(2)}\, + \,\tilde{\chi}_L^{(4)}$
$ \, + \, \cdots \, +\, \tilde{\chi}_L^{(2n)}$
is also a diagonal of a rational function~\cite{Short}, therefore 
this finite sum reduces to algebraic functions modulo powers
of primes, and, thus, satisfies functional equations modulo powers
of primes. Therefore $\, \tilde{\chi}_L$ which cannot
be distinguished from this sum modulo some $\, 2^r$ 
satisfies a functional equation  modulo some $\, 2^r$. 

 \vskip .1cm 

The question whether the full susceptibility satisfies
(probably involved) functional equations modulo primes 
different from $\, 2$, or powers of primes different 
from $\, 2^r$, remains open (even if, given (\ref{ifactor}) and 
(\ref{ifactor2}), it may seem unlikely).

 \vskip .1cm 

\vskip .1cm 

 \vskip .1cm 

Much remains to be done to understand, and describe,
this class of ``nice'' non-holonomic functions. It is worth recalling that,
while the product of two holonomic functions is a holonomic function, the 
{\em ratio of two holonomic functions} is, in general, {\em non-holonomic.} The class of 
functions that are expressible as {\em ratios of diagonals of rational functions,}
is clearly a very interesting and important class of functions: they 
are such that their series (i)
{\em can be recast into series with integer coefficients}, and (ii)
{\em modulo primes, or modulo powers of primes, reduce to algebraic 
functions}\footnote[1]{More generally, algebraic expressions of diagonal
of rational functions are such that they reduce modulo primes, 
or modulo power of primes, to algebraic functions.}. Concerning 
the susceptibility of the Ising model,
it is worth recalling that ratios of holonomic functions
can also yield\footnote[5]{The fact that solutions of a particular Painlev\'e-like 
non-linear third order differential equations, the Chazy III 
equations~\cite{Chazy1,Chazy2}, 
have (circular) natural boundaries is a {\em direct consequence} of the fact that 
the solutions correspond to the ratio of two holonomic functions, as shown 
by Chazy.
} a {\em natural boundary}. The ratio of diagonals of rational functions is probably 
an overly-simple scenario for the susceptibilitity of the Ising model. However it is
clearly important to start studying this class of functions, and further, to study
algebraic expressions of diagonals of rational functions, {\em per se}, before 
introducing them as a well-suited and powerful framework in which to study models of 
lattice statistical mechanics or enumerative combinatorics~\cite{Bousquet}.

 \vskip .1cm 

 \vskip .4cm 

 \vskip .1cm 

{\bf Acknowledgments:} One of us (JMM) would like to thank
 G. Christol for fruitful discussions on diagonals of rational functions,
 and discrete finite automata. A. J. Guttmann would like to thank the LPTMC 
for kind support, and gratefully acknowledges support for the 
Australian Research Council through grant DP140101110.
This work has been performed without
 any support of the ANR, the ERC, the MAE or any PES of the CNRS. 

 \vskip .2cm 

 \vskip .5cm 

\appendix

\section{Appendix: full susceptibility expansions in $\, w$}
\label{appendix}

\subsection{Low-temperature expansion in $\, w$}
\label{appendixlow}

Let us consider  (\ref{tildechilow}), the 
 low-temperature expansion for $\, \tilde{\chi}_L$ 
in the $\, w= \, \frac{1}{2}s/(1+s^2)$ variable
and introduce the series $\, {\tilde F}(w)\, = \, \, \tilde{\chi}_L/4$: 
\begin{eqnarray}
\label{tildeFlow}
\hspace{-0.95in}&&   \quad  \quad 
{\tilde F}(w)\, = \, \, 
{{\tilde{\chi}_L} \over {4}}    \,\, = \, \, \,\, 
\,{w}^{4} \, \,  +20\,{w}^{6}\,+350\,{w}^{8}
\,+5880\,{w}^{10}\,+97020 \,{w}^{12}\, +1585584\,{w}^{14}
\nonumber \\
\hspace{-0.95in}&&   \quad \quad \quad  \quad  \quad  \quad  \quad 
\,+25765744\,{w}^{16}\,+417159856\,{w}^{18}
\,\,\, + \,\,\, \cdots 
\end{eqnarray}
Modulo $2$ and $4$ the series (\ref{tildeFlow}) becomes simple polynomials:
\begin{eqnarray}
\label{tildeFlow24}
\hspace{-0.95in}&&    \quad \quad \quad  \,  \, 
  {\tilde F}_2(w)\, = \, \,w^4  \quad \,  \,  \,  \,  mod. \, \, 2,
\quad  \quad \quad   \,  \,  \, 
{\tilde F}_4(w)\, = \, \,w^4 \, +2\, w^8 \quad \,  \,  \, \,   mod. \, \, 4.
\end{eqnarray}

Modulo 8, this series becomes the lacunary series
\begin{eqnarray}
\label{tildeFlow8}
\hspace{-0.95in}&&  {\tilde F}_8(w)\, = \, \, 
{w}^{4} \, +4\,{w}^{6} \, +6\,{w}^{8} \, +4\,{w}^{12}  
+4\,{w}^{20}  +4\,{w}^{36}  +4\,{w}^{68}  +4\,{w}^{132}  +4\,{w}^{260}
\, + \,\, \cdots
\nonumber 
\end{eqnarray}
which satisfies the functional equation modulo 8:
\begin{eqnarray}
\label{funcequtildechilow2}
\hspace{-0.95in}&&   \quad  \quad 
{\tilde F}_8(w^2)  \,\, \,  - w^4 \cdot \, {\tilde F}_8(w) 
 \,  \, + \, 2 \, w^{10} \cdot \, (2 \, +w^2\,-w^6)
\, \, \, = \, \, \, \, 0  \quad \, \,\,  \,  \, \,  \,  mod. \, \, 8.
\end{eqnarray}
Comparing these results with $\, \tilde{\chi}_L^{(2)}/4$, 
the series expansion (\ref{tildechilow2})
divided by $\, 4$
\begin{eqnarray}
\label{tildechilow2over4}
\hspace{-0.95in}&&   \quad  \quad 
{{\tilde{\chi}_L^{(2)}} \over {4}}  \,\, = \, \, \,
 \, w^4 \cdot \,
 _2F_1\Bigl([{{3} \over{2}}, \,{{5} \over{2}}],[3], \, 16 \, w^2   \Bigr)   
\nonumber \\
\hspace{-0.95in}&&   \quad  \quad   \quad \quad
 \,\, = \, \, \, \,{w}^{4} \, \, \,  +20\,{w}^{6} \,\, \,   +350\,{w}^{8} \,\,
  +5880\,{w}^{10} \, \, +97020\,{w}^{12} \,\,  +1585584\,{w}^{14} 
\nonumber \\
\hspace{-0.95in}&&   \quad \quad \quad   \quad \quad  \quad 
+25765740\,{w}^{16}  \, +417159600\,{w}^{18} \, 
+6737127540\,{w}^{20} \,  \, \, +  \,  \, \cdots, 
\end{eqnarray}
one finds that this series (\ref{tildechilow2over4}) gives,
modulo $\, 2, \, 4, \, 8$, the same 
series expansions as (\ref{tildeFlow24}) and (\ref{tildeFlow8}), 
and consequently satisfies the {\em same functional equation
as} (\ref{funcequtildechilow2}). Again, similarly to the 
results displayed in section (\ref{avallowtemp}), in the variable 
$\, v$, one cannot make, modulo $\, 2, \, 4, \, 8$, a distinction,
for low-temperature expansions, between 
$\,\, \tilde{\chi}_L$  and $\,\, \tilde{\chi}_L^{(2)}$.

\vskip .1cm 

\subsection{High-temperature expansion in $\, w$}
\label{appendixhigh}

Let us consider (\ref{tildechi}), the  high-temperature expansion for 
$\, \tilde{\chi}_H$ in the $\, w= \, \frac{1}{2}s/(1+s^2)$ variable, 
and introduce the series $\, {\tilde F}(w)\, = \, \, \tilde{\chi}_H/2$:
\begin{eqnarray}
\label{tildeFapp}
\hspace{-0.95in}&&   \quad 
{\tilde F}(w)  \,\, = \, \, \, 
{{\tilde{\chi}_H} \over {2}}    \,\, = \, \, \,\,  \, 
w\, \, \, +4\,{w}^{2}\,\, +16\,{w}^{3}\,\, +64\,{w}^{4}\, +256\,{w}^{5}\, 
+1024\,{w}^{6}\, +4096\,{w}^{7}
\nonumber \\
\hspace{-0.95in}&&  \quad \quad  \quad   \quad  \quad  \quad \, 
+16384\,{w}^{8}\, +65540\,{w}^{9}\, +262144\,{w}^{10}
\, +1048720\,{w}^{11}\,\, + \, \, \cdots 
\end{eqnarray}
This series modulo $\, 2$ and $\, 4$ reads:
\begin{eqnarray}
\label{tildeF8app}
\hspace{-0.95in}&& \quad   \quad  \quad  \quad \, 
{\tilde F}_2(w)     \,\, = \, \, \, w 
\qquad mod. \, \, 2, 
\quad \quad \quad  \,  \, \, \, 
{\tilde F}_4(w)     \,\, = \, \, \,  w 
\qquad mod. \, \, 4, 
\end{eqnarray}
This series modulo 8 reads:
\begin{eqnarray}
\label{tildeF8app}
\hspace{-0.95in}&&   \,   \quad \quad \quad \quad 
{\tilde F}_8(w)     \,\, = \, \, \, \, 
w\, \, \, +4\,{w}^{2}\, +4\,{w}^{9}\, +4\,{w}^{17}\, 
+4\,{w}^{33}\, +4\,{w}^{65}\, +4\,{w}^{129}
\nonumber \\
\hspace{-0.95in}&&   \quad   \quad  \quad \quad \quad  \quad  \quad \quad \quad  
\, +4\,{w}^{257}\, +4\,{w}^{513}\,\, \, + \, \, \cdots
 \qquad mod. \, \, 8, 
\nonumber 
\end{eqnarray}
which satisfies the functional equation:
\begin{eqnarray}
\label{tildeF8func}
\hspace{-0.95in}&&   \quad  \quad \quad  \,  \,  \, 
{\tilde F}_8(w^2)\,\, + \,  4\,{w}^{3} \cdot \, ({w}^{7}-w+1) 
\,\, = \, \, \, w \cdot \, {\tilde F}_8(w) \quad \, \,  \, \,  \, \, mod. \, \, 8.
\end{eqnarray}
Modulo 16 it reads:
\begin{eqnarray}
\label{tildeF16app}
\hspace{-0.95in}&&   \quad \quad \quad \quad 
{\tilde F}_{16}(w)     \,\, = \, \, \,\,
w \,\, +4\,{w}^{2}\,\, +4\,{w}^{9}\,\, 
+12\,{w}^{17}\,\, +12\,{w}^{33}\,\, +12\,{w}^{65}
\nonumber \\
\hspace{-0.95in}&&   \quad   \quad  \quad \quad \quad \quad  \quad \quad 
\,+12\,{w}^{129}\,+12\,{w}^{257}\,+12\,{w}^{513}
\,\,\, + \,\, \cdots \qquad mod. \, \, 16, 
\end{eqnarray}
from which one deduces the functional relation:
\begin{eqnarray}
\label{tildeF16funcapp}
\hspace{-0.95in}&&   \quad  \, \,\, 
{\tilde F}_{16}(w^2)  \,  \, 
+ \, \, 4 \,{w}^{3} \cdot \, (2\,{w}^{15} +{w}^{7} -w +1)
  \,\,  \, = \, \, \,  \,
 w \cdot \, {\tilde F}_{16}(w) \qquad mod. \, \, 16.
\end{eqnarray}
Comparing the  series  (\ref{tildeFapp}) with the 
series $\, \tilde{\chi}_H^{(1)}/2$,
namely  the series (\ref{tildechi1}) divided by $\, 2$
\begin{eqnarray}
\label{tildechi1app}
\hspace{-0.95in}&&   
 {{\tilde{\chi}_H^{(1)}} \over {2}}
 \,\, = \, \, \, {{ \,w} \over {1\,-4\,w }} 
  \,\, = \, \,  \, \,  \,w\, \, +4\,{w}^{2}\, \, +16\,{w}^{3}\, \, 
+64\,{w}^{4}\, \, +256\,{w}^{5}\, \,  +1024\,{w}^{6}\,
 \nonumber \\
\hspace{-0.95in}&&   \quad  \, \, \,   
\,+4096\,{w}^{7}\, +16384\,{w}^{8}\,
+65536\,{w}^{9}\,\,
+262144\,{w}^{10}\,+ \, 1048576\,{w}^{11}\,\, + \,\, \cdots
\end{eqnarray}
one gets respectively, mod. $\, 2, \, 4, \, 8, \, 16, \, 32$: 
\begin{eqnarray}
\label{tildechi1app}
\hspace{-0.95in}&&   \quad 
w \quad mod. \, \, 2, \,\, \qquad \, \,  \, 
w \quad mod. \, \, 4, 
\qquad \,\,\, 
w \, + 4 \, w^2 \quad  mod. \, \, 8, 
\nonumber \\
 \hspace{-0.95in}&&   \quad 
w \, + 4 \, w^2 \, + 16 \, w^3  \quad \,\,\,  mod. \, \, 16, 
\qquad \quad 
w \, + 4 \, w^2 \, + 16 \, w^3  \quad \,\,\,  mod. \, \, 32, 
\qquad
 \\
 \hspace{-0.95in}&&   \quad 
w \, + 4 \, w^2 \, + 16 \, w^3  \quad \,\,\,  mod. \, \, 64, 
\qquad \quad 
w \, + 4 \, w^2 \, + 16 \, w^3 \, + 64 \, w^4 
 \quad \,\,\,  mod. \, \, 128.
\nonumber 
\end{eqnarray}

 \vskip .1cm 

\section{A very simple algebraic function example 
illustrating the emergence of a lacunary series}
\label{heuristic}

 Let us consider a very simple algebraic function, 
 the {\em Catalan number generating function}~\cite{Kauers}:
\begin{eqnarray}
\label{Catalan}
\hspace{-0.95in}&&   \quad \quad \quad \quad \quad  \quad  \quad  
 C(x)\,\,\,\, = \,\,\,\, \,
{\frac {1-\sqrt {1\, -4\,{x}}}{2 \, x}}.
\end{eqnarray}
It is the solution of the quadratic equation 
$\, x \cdot \, C(x)^2 \, -C(x) \, +1\, = \, \, 0$.
Modulo 2 the series $\, x \cdot \, C(x)$ reduces 
to $\, L(x)\, -1$,  where $\, L(x)$ is the {\em lacunary} series : 
\begin{eqnarray}
\label{Lx}
\hspace{-0.95in}&&   \quad \,\,  \quad 
L(x) \, \, = \,\,\,\,\, 
1\,\, +\,x\, \,+\,{x}^{2}\, +\,{x}^{4}\, +\,{x}^{8}\, +\,{x}^{16}\,
 +\,{x}^{32} \, +\,{x}^{64}\, +\,{x}^{128}\, +\,{x}^{256}\, 
\nonumber \\
\hspace{-0.95in}&&   \quad \quad 
\quad \quad \quad \quad \quad   \quad \, \, \, \,
+\,{x}^{512}\, +\,{x}^{1024}\,\, +\,{x}^{2048} \, 
\,\,   + \, \, \, \cdots
\end{eqnarray}
This Catalan number generating function (\ref{Catalan}) satisfies 
the {\em functional equation} 
\begin{eqnarray}
\label{C2x}
\hspace{-0.95in}&&   \quad \,\,  \quad \quad \quad   \quad   \quad 
x \cdot \, C(x^2) \, \, -C(x)\,  \, +1  \,
\, \, = \, \, \, 0  \quad \quad \,\, \, \, mod. \, \, 2. 
\end{eqnarray}
which can be seen as the consequence of 
$\,  C(x^2)\, = \, \, C(x)^2 \quad mod. \, \, 2$, 
or as the consequence of the functional equation
 $\, \,  L(x^2)\, = \, \, L(x) \, + \, x$.

This generating function (\ref{Catalan}) yields many other 
lacunary series modulo $\, 2^r$ (see for instance~\cite{Kauers}), 
for instance, modulo $\,8$ the series expansion of
 $\, 4+4/(1 \, -x\cdot \, C(x^2))$ reduces to $\, 4 \cdot \, L(x)$ 
where $\, L(x)$ is given by (\ref{Lx}). This result, namely 
the emergence of lacunary series, can be seen as a simple example 
of the previous finite automaton 
results, or, equivalently, congruence on algebraic functions, 
in the simplest case where only square roots occur.

\end{document}